\documentclass{aa}
 \usepackage{natbib}

\usepackage{amssymb,alltt}
\usepackage{latexsym}
\usepackage{graphicx}
\usepackage[english]{babel}
\usepackage{indentfirst}

\vspace{1cm}
\def\R{\mathbb{R}}
\def\T{\mathbb{T}}

\def\Z{\mathbb{Z}}
\def\P{\mathbb{P}}
\def\Q{\mathbb{Q}}

\def\u{{\bf u}}
\def\v{{\bf v}}
\def\w{{\bf w}}

\def\k{{\bf k}}
\def\n{{\bf n}}

\usepackage[intlimits]{amsmath}
 \usepackage{amsfonts,upgreek,bbm,mathrsfs}
 \usepackage{amssymb,alltt}
 \usepackage{latexsym} 
\usepackage[latin1]{inputenc} 
\usepackage{abbrevjournaux}
\usepackage{longtable,multirow,enumerate,threeparttable}
 \usepackage{lineno,ulem,cancel}
\usepackage{theorem,algorithm,algorithmic}
\usepackage[ruled,vlined,algo2e]{algorithm2e}
\usepackage{latexsym}
\usepackage{amsmath}
\usepackage{amssymb}
\usepackage{amsfonts}
\usepackage{amsbsy}
\usepackage{graphics}
\usepackage{euscript}
\usepackage{varioref}
\usepackage{enumerate}
\usepackage{verbatim}
\usepackage{makeidx}
\usepackage{color}

\begin{document}
\title{Wavelet Helmholtz decomposition for weak lensing mass map reconstruction}
 
\author{E. Deriaz \inst{1}, J.-L. Starck \inst{2}, and S. Pires \inst{2}}
\institute{Laboratoire de M\'ecanique Mod\'elisation et Proc\'ed\'es Propres - UMR-6181 CNRS, IMT La Jetée,
Technopôle de Château-Gombert,
38, Rue Frédéric Joliot-Curie,
13451 MARSEILLE Cedex 20, France
\and
Laboratoire AIM, UMR CEA-CNRS-Paris 7, Irfu, SEDI-SAP, Service d'Astrophysique, CEA Saclay,
F-91191 GIF-Sur-YVETTE CEDEX, France }
 

\offprints{erwan.deriaz@l3m.univ-mrs.fr}
 
\date{\today}

 

\abstract{{To derive the convergence field from the gravitational shear $\gamma$ of the background galaxy images, the classical methods require a convolution of the shear to be performed over the entire sky, usually expressed thanks to the Fast Fourier transform (FFT)}. However, it is not optimal for an imperfect geometry survey. Furthermore, FFT implicitly uses periodic conditions that introduce errors to the reconstruction. {A method has been proposed that relies on computation of an intermediate field $u$ that combines the derivatives of $\gamma$ and on convolution with a Green kernel.}
In this paper, we study the wavelet Helmholtz decomposition as a new approach to reconstructing the dark matter mass map. We show that a link exists between the Helmholtz decomposition and the E/B mode separation. 
We introduce a new wavelet construction, that has a property that gives us more flexibility in handling the border problem, and we propose a new method of reconstructing the dark matter mass map in the wavelet space.
A set of experiments {based on noise-free images} illustrates that this Wavelet Helmholtz decomposition reconstructs the borders better than all other existing methods.}

\maketitle 
\markboth{Wavelet Helmholtz decomposition for Weak Lensing mass map reconstruction}{}

\keywords{Cosmology : Weak Lensing, Methods : Helmholtz decomposition, Multi-scale Analysis}

\section{Introduction}

Weak gravitational lensing provides a method of directly mapping the distribution of dark matter in the universe \citep{wlens:bartelmann99,wlens:mellier99,wlens:vanwaerbeke01,wlens:mellier02,wlens:refregier03,wlens:munshi08,wlens:pires10}. This method is based on the weak distortions that lensing induces in the images of background galaxies as it travels toward us through intervening structures. The induced shear is small, typically around an order of magnitude less than the ellipticity seen in background galaxies, therefore the weak lensing effect cannot be measured on a single galaxy. To measure the shear field, it is necessary to measure ellipticities of many background galaxies and construct a statistical estimate of their systematic alignment. This measured shear field is directly related to the distribution of the (dark) matter and thus to the geometry and the dynamics of the universe. As a consequence, weak gravitational lensing offers unique possibilities for probing the statistical properties of dark matter and dark energy in the universe.

The mapping of the dark matter distribution has become a central topic in weak lensing since the very first reconstructed 2D mass maps demonstrated that you can see the dark components of the universe. Recently, this field has seen great success in reconstructing the deepest and largest 2D dark matter distribution in the COSMOS field \citep{wlens:massey07}. From this distribution, the nature of dark matter can be better understood, so better constraints can be set on dark energy {\citep{schrabback10}}. 
This method is now widely used but, the amplitude of the weak lensing signal is so weak that its detection relies on the accuracy of the techniques used to analyze the data. Each step in the analysis has required the development of advanced techniques dedicated to these applications.

The reconstruction of the dark matter mass map from shear measurements is a difficult inverse problem because of observational effects such as noise and the complex geometry of the field. Classical methods based on \citet{wlens:kaiser93} are not optimal. Indeed they require a convolution of the observed shear with a kernel to be performed over the entire field, usually expressed in the Fourier domain, and the FFT imposes a condition of periodicity that is not true and introduces a mixing between E and B modes. 
An alternative interesting method was proposed in \citet{wlens:seitz96}, which consists in convolving the shear field with a local kernel that depends on the shape of the domain. 
It was shown to be optimal in \citet{wlens:lombardi98}, and it was reformulated later in \citet{seitz2001}.
 
This approach is relatively expensive in computational resources. Much effort has been made in past years to properly derive a cosmic shear two-point correlation on a finite interval \citep{wlens:kilbinger06,wlens:schneider07,wlens:fu2010}. In \citet{wlens:fu2010}, 
a new statistic is proposed, maximizing the signal-to-noise ratio and a figure of merit based on the Fisher matrix 
of the cosmological parameters $\Omega_m$ and $\sigma_8$. However, this approach does not allow reconstructing a map, and cannot be used for higher order statistics studies such as bispectrum analysis \citep{wlens:vanwaerbeke01,starck:pires08,munshi11}, {peak counting \citep{marian11}, 
Minkowski Functionals \citep{kratochvil11}},  or wavelet peak counting \citep{pires09}.

In this paper, we study a new approach for weak lensing mass inversion in the ideal case of noise free data based on a Helmholtz decomposition in the wavelet domain. 
We first show that there is an explicit relation between the weak lensing E/B mode decomposition and the Helmholtz decomposition which decomposes a vector field into two components, one curl-free and the second divergence-free. In contrast to the method presented in \citet{wlens:seitz96}, our method 
does not rely on a convolution with a kernel, and no intermediate field $\u$ derived from the shear field is needed.
The Helmholtz decomposition is performed in the wavelet space, and the Fourier transform is never used. This gives us more flexibility to handle borders or to add some constraints in our solution. We introduced a new wavelet construction based on border-wavelets, and 
we present two kinds of wavelet Helmholtz decomposition algorithms. 
The first one imposes a global zero B modes constraint and the second a local zero B mode constraint only on the borders. 
 
This paper is structured as follows. In Sect. 2, we review the basis of the weak lensing formalism. In Sect. 3, we introduce of the Helmholtz decomposition to the weak lensing E/B modes decomposition, {and the discretization is discussed in Sect. 4.
Then, in Sect. 5, the Helmholtz decomposition is extended to the wavelet domain and, we present new constructions of divergence-free and curl-free wavelets that satisfy suitable boundary conditions. And finally, in Sect. 6, we do some numerical experiments for weak lensing and conclude in Sect. 7. In Appendix A, we provide the mathematical expressions for the vector wavelets and in Appendix B, we present the algorithm to perform the E mode convergence reconstruction in a bounded domain using a wavelet Helmholtz decomposition.}

\section{Weak lensing formalism}
 \subsection{The weak lensing equations} 
 
In weak lensing surveys, the observable shear field $\boldsymbol\gamma_i$ can be written in terms of the lensing gravitational potential $\psi(\theta_i)$ \citep{wlens:bartelmann99}:
\begin{equation}
\label{gamma}
\begin{array}{l} \gamma_1= \frac{1}{2}(\partial_1^2-\partial_2^2)\psi\\ \gamma_2=\partial_1\partial_2\psi, \end{array}
\end{equation}
where the partial derivatives $\partial_i$ are with respect to $\theta_i$.
Notice the difference in sign with respect to \citep{wlens:seitz96,wlens:lombardi98}.

The shear $\gamma_i(\theta)$ with $i = 1, 2$ is derived from the shapes of galaxies at positions $\theta$ in the image. The $\gamma_1$ component corresponds to a deformation in the horizontal/vertical direction and the $\gamma_2$ component to a deformation in the diagonal direction (see Fig. \ref{gamma_defo}).

The convergence $\kappa(\theta)$ can also be expressed in terms of the lensing potential $\psi$,
\begin{equation}
\label{kappa}
\kappa=\frac{1}{2}(\partial_1^2 + \partial_2^2) \psi,
\end{equation}
and is related to the mass density $\Sigma(\theta)$ projected along the line of sight by
\begin{equation}
\label{sigma1}
\kappa(\theta) = \frac{\Sigma(\theta)}{\Sigma_{crit}},
\end{equation}
where the critical mass density $\Sigma_{crit}$ is given by
\begin{equation}
\label{sigma2}
\Sigma_{crit}=\frac{c^2}{4\pi G}\frac{D_{OS}}{D_{OL}D_{LS}},
\end{equation} 
where $G$ is Newton's constant, $c$ the speed of light, and $D_{OS}$, $D_{OL}$, and $D_{LS}$ are respectively the angular-diameter distances between the observer and the galaxies, the observer and the lens, and the lens and the galaxies.

{
\subsection{Difference between the sheared galaxies and the shear field $\boldsymbol \gamma$}

The deformation of a circle $\mathcal{C}$ by a shear field provides an ellipse $M\mathcal{C}$ and corresponds to the local linear application of the matrix
\begin{equation}
M=(1-\kappa)\,\left(\begin{array}{cc} 1 & 0 \\ 0 & 1 \end{array} \right)+|\gamma|\,\left(\begin{array}{cc} \cos\,2\phi & \sin\,2\phi \\ \sin\,2\phi & \cos\,2\phi \end{array} \right),
\end{equation}
where $\kappa$ is the convergence, $\phi$ the angle formed by the larger axis of the ellipse and the $(Ox)$ axis, and $|\gamma|$ the amplitude of the deformation.

\begin{figure}[htb]
\begin{center}
\input{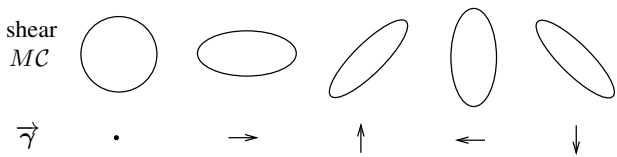}
\end{center}
\caption{Galaxy shapes and corresponding gravitational shears $\boldsymbol \gamma$ for a circular galaxy.}
\label{gamma_defo}
\end{figure}

In the weak lensing approximation, we only consider the shear $\boldsymbol\gamma=|\gamma|(\cos\,2\phi,\sin\,2\phi)$ which concentrates the information on the angle of deformation and the amplitude of deformation and defines a 2D vector field: $\boldsymbol\gamma:\Omega \to \R^2$, see figure \ref{gamma_defo}. Although the shear is often represented by sheared galaxies which is a spin-2 representation, as one can see in the first row of Fig. \ref{gamma_defo}, the shear field $\boldsymbol \gamma$ itself is a usual vector field, as seen in the second row of Fig. \ref{gamma_defo}, so we can compute its divergence and its curl:
\begin{equation}
\label{div-curl}
{\rm div}\, \boldsymbol \gamma=\partial_1 \gamma_1+\partial_2 \gamma_2,
\quad {\rm curl}\, \boldsymbol \gamma=\partial_1 \gamma_2-\partial_2 \gamma_1,
\end{equation}
and apply the Helmholtz decomposition to it.
}

\subsection{The mass inversion}
{The weak lensing mass inversion problem consists in reconstructing the projected (normalized) mass distribution $\kappa(\theta)$ from the measured shear field $\gamma_i(\theta)$. We invert Eq. (\ref{gamma}) to find the lensing potential $\psi$ and then apply formula (\ref{kappa}) following the classical methods based on the pioneering work of \citet{wlens:kaiser93}. In short, this corresponds to \citep{wlens:kaiser93,wlens:starck06}:}
\begin{eqnarray}
\label{eqn_reckE}
\tilde \kappa  & = & \Delta^{-1}\left((\partial_1^2 - \partial_2^2) \gamma_1+ 2 \partial_1\partial_2 \gamma_2\right)  \nonumber  \\
   & =  & \frac{\partial_1^2 - \partial_2^2}{\partial_1^2 + \partial_2^2} \gamma_1+ \frac{2 \partial_1\partial_2}{\partial_1^2 + \partial_2^2}\gamma_2.
\end{eqnarray}
This method usually is called a Fourier transform and has been generalized to spherical harmonics in \citet{wlens:pichon09} to deal with full-sky maps. To recover $\kappa$ from both $\gamma_1$ and $\gamma_2$ there is a degeneracy when $k_1 = k_2 = 0$. Therefore, the mean value of $\kappa$ cannot be recovered from the shear maps. This is known as the mass-sheet degeneracy.

The most important drawback of this method is that it requires an observation of the shear over the entire sky. As a result, if the observed shear field has a finite size or a complex geometry, then the method can produce artifacts on the reconstructed convergence distribution near the boundaries of the observed field. The error caused by the artificial periodic boundary conditions can not be decreased by taking mirror conditions which remove the discontinuity.

There are local inversions that reduce the unwanted boundary effects \citep{wlens:seitz96}. But regardless of the local formula used, the reconstructed field has more noise than when obtained with a global inversion, and the inversion is unstable in the presence of uncorrelated noise. 

The convergence $\kappa$ can be computed directly in direct space (without the Fourier transform) thanks to the kernel integration \citep{wlens:seitz96}:

\begin{equation}
\kappa(\theta)-\kappa_0=\frac{1}{\pi}\int_{\theta' \in \Omega}K(\theta,\theta')\cdot {\boldsymbol \gamma}(\theta-\theta')\, d\theta',
\end{equation}
where $\kappa_0$ stands for the mean value of $\kappa$. The kernel $K$ depends on the geometry of the domain $\Omega$. For $\Omega=\R^2$, it is given by
\begin{equation}
K(\varphi,\theta)=\left(\frac{\theta_2^2-\theta_1^2}{(\theta_1^2+\theta_2^2)^2} , \frac{-2\theta_1\theta_2}{(\theta_1^2+\theta_2^2)^2}\right) .
\end{equation}
In \citet{wlens:seitz96}, the authors propose to combine the derivatives of ${(\gamma_i)}$
\begin{equation}
\u=\left(\begin{array}{l} \partial_1\gamma_1+\partial_2\gamma_2 \\ \partial_1\gamma_2-\partial_2\gamma_1 \end{array}\right),
\end{equation}
and then to apply the Helmholtz decomposition $\u=\nabla\kappa^{(E)}+\nabla\times\kappa^{(B)}$, in order to reconstruct the convergence $\kappa=\kappa^{(E)}$. 

In the following, we propose to use the Helmholtz decomposition directly on $(\gamma_i)$ to perform the E/B mode decomposition, and this
can be seen as a kind of shortcut in comparison with the Seitz \& Schneider method. Furthermore, we also propose to perform the 
Helmholtz decomposition in the wavelet space using a new kind of border-wavelet, which allows us to reduce border effects during the reconstruction.

\subsection{The E and B mode decomposition}

Just as a vector field can be decomposed into a gradient or electric (E) component, and a curl or magnetic (B) component, the shear field $\gamma_i({\mathbf \theta})$ can be decomposed into two components, which for convenience we also call (E) and (B). The decomposition of the shear field into each of these components can be easily performed by noticing that a pure E mode can be
transformed into a pure B mode by a clockwise rotation of the sheared galaxies by 45$^{\circ}$: $\gamma_{1} \rightarrow - \gamma_{2}$, $\gamma_{2}
\rightarrow \gamma_{1}$ (see fig.~\ref{gamma_defo}).

Because the weak lensing arises from a scalar potential (the Newtonian potential $\Phi$), it can be shown
that weak lensing only produces E modes. As a result, the least square estimator of the E modes convergence field is simply
\begin{equation}
\label{kappaE}
\tilde \kappa_{E} = \Delta^{-1}\left((\partial_1^2 - \partial_2^2) \gamma_1+ 2 \partial_1\partial_2\gamma_2\right).
\end{equation}
The E mode deformations $\boldsymbol \gamma^{(E)}$ which satisfy (\ref{gamma}) can be equivalently defined by
\begin{equation}
\label{GammaE}
\boldsymbol \gamma^{(E)}\in \Gamma_E=\left\{ \boldsymbol \gamma ~:~2\partial_1\partial_2\gamma_1-(\partial_1^2-\partial_2^2)\gamma_2=0 \right\}.
\end{equation}

On the other hand, residual systematics arising from imperfect correction of the instrumental PSF
or telescope aberrations, generally generates both E and B modes. The presence of B modes can thus be used to test for any residual systematic effects in current weak lensing surveys. 
For this purpose, the following estimator for the B mode convergence field can be formed \citep{wlens:starck06},
\begin{equation}
\label{kappaB}
\tilde \kappa_{B} = \Delta^{-1}\left(2 \partial_1\partial_2\gamma_1-(\partial_1^2 - \partial_2^2) \gamma_2\right),
\end{equation}
and it is consistency with zero will confirm the absence of systematics.\\

The B mode deformations $\gamma^{(B)}$ derive from potentials $\psi_B$ as
\begin{equation}
\label{B mode}
\begin{array}{l} \gamma_{1}^{(B)}=\partial_1\partial_2\psi_B, \\ \gamma_{2}^{(B)}=-\frac{1}{2}(\partial_1^2-\partial_2^2)\psi_B, \end{array}
\end{equation}
and form the set
\begin{equation}
\label{GammaB}
\boldsymbol \gamma^{(B)}\in \Gamma_B=\left\{ \boldsymbol \gamma ~:~(\partial_1^2-\partial_2^2)\gamma_1+2\partial_1\partial_2\gamma_2=0 \right\}.
\end{equation}
If we consider $\boldsymbol \gamma$ defined on the infinite plan $\R^2$ or on the torus $\T^2$, i.e.
$\boldsymbol \gamma\in (L^2(\R^2))^2$ or $\boldsymbol \gamma\in (L^2(\T^2))^2$, then $\Gamma_E$ and $\Gamma_B$
define orthogonal spaces for the usual scalar product in $L^2$. This is no longer the case for domains with boundaries.
In this case, $\Gamma_E$ and $\Gamma_B$ have nonzero common elements.

Solutions (\ref{kappaE}) and (\ref{kappaB}) are therefore no longer valid for bounded domains.
The gravitational shear that derives from the bi-harmonic function $\psi$ such that $\Delta^2\psi=0$ produces modes that both remain to $\Gamma_E$ and $\Gamma_B$ \citep{wlens:bunn03}.
Indeed, the borders cause mode-mixing or, equivalently, a leakage of E modes into B modes. Therefore, even a pure gravitational shear will produce both E and B modes in a finite field. Consequently, it will be more difficult to confirm the absence of residual systematics with the presence of border effects.
More details can be found in \citet{crittenden2002,schneider2002}.

In the next section, we propose to explore the use of the Helmholtz decomposition to reduce boundary effects in E and B mode decomposition.

\section{Introduction of the Helmholtz decomposition into the weak lensing formalism}
\label{Helm_E/B}

In this section, we introduce the relation between the Helmholtz decomposition and the weak lensing E/B mode decomposition. 
The Helmholtz decomposition in two orthogonal components only stands for unbounded domains. Then the unicity of the $\kappa_{E}$ reconstruction on bounded domains can only be obtained under some constraints. 

\subsection{The weak lensing E/B mode decomposition}
\label{EBmodes}

On $(L^2(\R^2))^2$, we can express the relations (\ref{gamma}) \& (\ref{B mode}) thanks to the Fourier transform:
\begin{eqnarray}
\hat{\boldsymbol \gamma}^{\bf (E)}(\boldsymbol k) & = & -\left[\begin{array}{c} \frac{1}{2} (k_1^2- k_2^2) \\ k_1 k_2 \end{array}\right]\hat\psi^{\bf (E)}(\boldsymbol k),\quad \nonumber \\
\hat{\boldsymbol \gamma}^{\bf (B)}(\boldsymbol k) & = & -\left[\begin{array}{c} k_1 k_2 \\ \frac{1}{2} (- k_1^2+ k_2^2) \end{array}\right]\hat\psi^{\bf (B)}(\boldsymbol k).
\end{eqnarray}
It allows us to derive the convergence $\kappa_{E}$ deriving from the E modes (the \emph{actual} convergence)
and the convergence $\kappa_{B}$ deriving from the B modes (the \emph{systematic errors}) similarly to
(\ref{kappaE}) and (\ref{kappaB}):
\begin{equation}
\label{kappa12}
\hat{\boldsymbol\kappa}(\boldsymbol k)=\left[ \begin{array}{c} \hat\kappa_{E}(\boldsymbol k) \\ \hat\kappa_{B}(\boldsymbol k) \end{array}\right]
=\underbrace{\frac{1}{|\boldsymbol k|^2}\left[ \begin{array}{cc} k_1^2- k_2^2 & 2 k_1 k_2 \\ 2 k_1 k_2 & - k_1^2+ k_2^2 \end{array}\right]}_{A_{\boldsymbol\kappa}}
\left[ \begin{array}{c} \hat\gamma_1(\boldsymbol k) \\ \hat\gamma_2(\boldsymbol k) \end{array} \right],
\end{equation}
where hat symbols denote the Fourier transforms.\\


\subsection{Relation between Helmholtz and E/B modes decompositions on an unbounded domain}

The Helmholtz decomposition theorem says that every vector field ${\boldsymbol\gamma}$, defined everywhere in space
can be decomposed into a rotational part ${\boldsymbol\gamma}_{\rm div\,0}$ and an irrotational part ${\boldsymbol\gamma}_{\rm curl\,0}$.
{The Helmholtz decomposition consists, for a vector field ${\boldsymbol\gamma}$, in writing
\begin{equation}
{\boldsymbol\gamma} = {\boldsymbol\gamma}_{\rm div\,0}+{\boldsymbol\gamma}_{\rm curl\,0}=\left( \begin{array}{c} \partial_2 \psi \\ -\partial_1 \psi \end{array} \right)+\left( \begin{array}{c} \partial_1 p \\ \partial_2 p \end{array} \right),
\label{Helmholtz1}
\end{equation}
 where ${\boldsymbol\gamma}_{\rm div\,0}=\nabla\times\psi$ and ${\boldsymbol\gamma}_{\rm curl\,0}=\nabla p$ with $\psi$ and $p$ differentiable functions.}

This decomposition is unique for domains without boundaries.
In the Fourier domain, it consists in decomposing the vector field $\hat{{\boldsymbol\gamma}}(\boldsymbol k)$ in the longitudinal direction, i.e. parallel to $\boldsymbol k$ and in the transverse direction, i.e. perpendicular to $\boldsymbol k$. The longitudinal component corresponds to the curl-free component and the transverse component to the divergence-free component.

Using the following notations: $\P{\boldsymbol\gamma}={\boldsymbol\gamma}_{\rm div\,0} $ and $\Q {\boldsymbol\gamma}={\boldsymbol\gamma}_{\rm curl\,0}$, we obtain
\begin{equation}
\hat {\boldsymbol\gamma}=\widehat{\P {\boldsymbol\gamma}} + \widehat{\Q {\boldsymbol\gamma}},
\label{Helmholtz2}
\end{equation}
with:
\begin{eqnarray}
\label{Pand}
\widehat{\P {\boldsymbol\gamma}}(\boldsymbol k) & = & \underbrace{\frac{1}{|\boldsymbol k|^2}
\left[ \begin{array}{cc} k_2^2 & - k_1 k_2 \\ - k_1 k_2 & k_1^2 \end{array}\right]}_{A_{\P}}
\hat{\boldsymbol\gamma}(\boldsymbol k)
\quad
\\
\label{andQ}
\quad
\widehat{\Q {\boldsymbol\gamma}}(\boldsymbol k) & = & \underbrace{\frac{1}{|\boldsymbol k|^2}
\left[ \begin{array}{cc} k_1^2 & k_1 k_2 \\ k_1 k_2 & k_2^2 \end{array}\right]}_{A_{\Q}}
\hat{\boldsymbol\gamma}(\boldsymbol k).
\end{eqnarray}
From equations (\ref{kappa12}), (\ref{Pand}), and (\ref{andQ}), we have $A_{\boldsymbol\kappa}=-A_{\P}+A_{\Q}$.
Then
\begin{equation}
\boldsymbol\kappa = \left[ \begin{array}{c} \hat\kappa_{E}(\boldsymbol k) \\ \hat\kappa_{B}(\boldsymbol k) \end{array}\right] = 
-A_{\P} \left[ \begin{array}{c} \hat\gamma_1(\boldsymbol k) \\ \hat\gamma_2(\boldsymbol k) \end{array} \right] 
+A_{\Q} \left[ \begin{array}{c} \hat\gamma_1(\boldsymbol k) \\ \hat\gamma_2(\boldsymbol k) \end{array} \right].
\label{relation2}
\end{equation}

This provides the following E and B modes convergence:
\begin{eqnarray}
\label{relation4}
\ \quad {\rm if}\quad {\boldsymbol\gamma}&=&\P {\boldsymbol\gamma} + \Q {\boldsymbol\gamma},\\
\label{relation3}
\ \quad {\rm then}\quad
\boldsymbol\kappa&=&\left[ \begin{array}{c} \kappa_{E} \\ \kappa_{B}\end{array}\right] =-\P \boldsymbol\gamma+\Q \boldsymbol\gamma.
\end{eqnarray}

\subsection{Relation between Helmholtz and E/B mode decompositions on a bounded domain}

On the square $\Omega=(0,1)^2$, the Helmholtz decomposition becomes much more complicated, see \citep[][ for further details]{wlens:lions84}. Indeed, the unicity of the Helmholtz decomposition ${\boldsymbol\gamma}=\P {\boldsymbol\gamma}+ \Q {\boldsymbol\gamma}$ disappears. However, it can be regained by considering three components in the orthogonal decomposition:
\begin{equation}
{\boldsymbol\gamma}= {\boldsymbol\gamma}_{\rm div\,0} + {\boldsymbol\gamma}_{\rm curl\,0} + {\boldsymbol\gamma}_{\rm harm},
\end{equation}
with ${\rm div}\,{\boldsymbol\gamma}_{\rm div\,0}=0$ on the square $\Omega$ and ${\boldsymbol\gamma}_{\rm div\,0}\cdot{\bf n}=0$ on the borders of the square $\partial\Omega$,
${\rm curl}\,{\boldsymbol\gamma}_{\rm curl\,0}=0$ on $\Omega$ and ${\boldsymbol\gamma}_{\rm curl\,0}\cdot{\boldsymbol\tau}=0$ on $\partial\Omega$,
and ${\rm div}\,{\boldsymbol\gamma}_{\rm harm}={\rm curl}\,{\boldsymbol\gamma}_{\rm harm}=0$ on $\Omega$, where ${\bf n}$ and ${\boldsymbol\tau}$
are the normal and tangential unit vectors on $\partial \Omega$. This new term ${\boldsymbol\gamma}_{\rm harm}$ 
derives from a harmonic function on $\Omega$: ${\boldsymbol\gamma}_{\rm harm}= \nabla p_0 = \nabla \times \omega_0$ with
$\Delta p_0=\Delta \omega_0=0$. Alternatively we can write ${\boldsymbol\gamma}_{\rm harm}=\nabla p_0+\nabla \times \omega_0$ where $p_0$
and $\omega_0$ are harmonic functions.\\

Considering the Helmholtz decomposition of $ \boldsymbol\gamma=\P\boldsymbol\gamma+\Q\boldsymbol\gamma$,
the new term ${\boldsymbol\gamma}_{\rm harm}$ can be included either in the divergence-free part or in the curl-free part of
the decomposition. This decomposition is no longer unique, and it is the same for the E/B mode decomposition (\ref{relation3}).\\

However, the unicity of the reconstruction of $\kappa_{E}$ can be obtained under the condition $\kappa_{B}=0$.
Indeed, if $\boldsymbol\gamma\in (L^2([0,1]^2)^2$ is split into a divergence-free part $\P{\boldsymbol\gamma}$ and
a curl-free part $\Q{\boldsymbol\gamma}$, satisfying
\begin{equation}
\label{solHelmWL}
 \boldsymbol\gamma=\P\boldsymbol\gamma+\Q\boldsymbol\gamma \quad {\rm and} \quad
\left(\begin{array}{c} \kappa_{E} \\ 0 \end{array}\right)=-\P\boldsymbol\gamma+\Q\boldsymbol\gamma,
\end{equation}
then, $\boldsymbol\gamma$ remains to $\Gamma_E$ as defined in (\ref{GammaE}).
In addition, the convergence $\kappa_{E}$ is uniquely defined, modulo an additive constant.
Reciprocally, if $\boldsymbol\gamma$ remains to $\Gamma_E$, then we have the decomposition 
\begin{eqnarray}
\boldsymbol\gamma & = & \left(\begin{array}{c} \frac{\partial_1^2-\partial_2^2}{2} \\ \partial_1\partial_2 \end{array}\right)\psi \nonumber \\ 
 & = & \underbrace{\frac{1}{2}\left(\begin{array}{c} -\partial_2 \\ \partial_1 \end{array}\right)\partial_2 \psi}_{=\P\boldsymbol\gamma}
+\underbrace{\frac{1}{2}\left(\begin{array}{c} \partial_1 \\ \partial_2 \end{array}\right)\partial_1 \psi}_{=\Q\boldsymbol\gamma},
 \quad 
 \end{eqnarray}
and
\begin{eqnarray}
\boldsymbol\kappa = \left[\begin{array}{c} \kappa_E \\ \kappa_B \end{array}\right]
 = -\frac{1}{2}\left(\begin{array}{c} -\partial_2 \\ \partial_1 \end{array}\right)\partial_2 \psi
+\frac{1}{2}\left(\begin{array}{c} \partial_1 \\ \partial_2 \end{array}\right)\partial_1 \psi,
 \quad 
 \end{eqnarray}
so $\kappa_{B}=-\frac{1}{2}\partial_1\partial_2 \psi+\frac{1}{2}\partial_2\partial_1 \psi=0.$
 The same is true for the B modes. Under the constraint $\kappa_{E}=0$, $\boldsymbol\gamma$ remains to $\Gamma_B$ as defined in (\ref{GammaB}) and $\kappa_{B}$ is uniquely defined, modulo an additive constant.

This method offers a way to recover the E modes convergence without the border effects.
But the constraint $\kappa_{B}=0$ should be used carefully, because it assumes that there are no residual systematics in the data due to the observational or instrumental effects. We see in the following that we can relax this constraint, and impose that $\kappa_{B}=0$ only on the border, and have a good reconstruction of E modes, even in the presence of B modes.

\section{Discretization of the problem}

\subsection{Using a harmonic decomposition}
\label{minimisation}

To solve the weak lensing problem on a bounded domain $\Omega$ using the formula (\ref{solHelmWL}), a possible solution consists in writing $\boldsymbol \gamma$ as
\begin{equation}
\label{decomp/Omega}
\boldsymbol \gamma=\nabla p +\nabla \times \omega+\nabla p_0 +\nabla \times \omega_0,
\end{equation}
with
\begin{equation}
\label{divcurlpart}
\Delta p={\rm div}\,\boldsymbol \gamma, \quad \Delta \omega={\rm curl}\,\boldsymbol \gamma,\\
\end{equation}
where $p=0$ and $\omega=0$ on the borders of the square $\delta \Omega$ and
\begin{equation}
\label{harmpart}
 \Delta p_0=0, \quad \Delta \omega_0=0, \\
\end{equation}
i.e., $p_0$ and $\omega_0$ are harmonic functions on $\Omega$.

From the relations (\ref{relation3}) and (\ref{decomp/Omega}), we can rewrite the E/B modes decomposition as
\begin{equation}
\label{rela}
\boldsymbol\kappa=\left[ \begin{array}{c} \kappa_{E} \\ \kappa_{B}\end{array}\right] =-(\nabla p+\nabla p_0) + (\nabla \times \omega +\nabla \times \omega_0).
\end{equation}
This decomposition is not unique, and we have to find how to split the harmonic term ${\boldsymbol\gamma}_{\rm harm}$ into $\nabla p_0$ and $\nabla \times \omega_0$.

\ 

By assuming $\gamma$ derives from the gravitational potential without residual systematics, we can apply to the solution the constraint that $\kappa_B = 0$ on all the domain $\Omega$. 
Then, the solution to this problem becomes unique and the convergence is given by Eq. (\ref{rela}).

Then, $p_0$ and $\omega_0$ can be estimated accurately by minimizing the following quantity:
\begin{equation}
\label{approx_g}
\min_{p_0, \omega_0} \left\|\boldsymbol \gamma-\nabla p -\nabla \times \omega-\nabla p_0 -\nabla \times \omega_0\right\|^2 + \left\| \kappa_B \right\|^2,
\end{equation}
where $\nabla p$ and $\nabla \times \omega$ can be obtained from the relation (\ref{divcurlpart})
and $\nabla p_0$ and $\nabla \times \omega_0$ can be obtained by decomposing $p_0$ and $\omega_0$ in a basis of harmonic functions on $\Omega$ ($p_0=\sum_{n=1}^N c_n \psi_n$ and $\omega_0=\sum_{n=1}^N d_n \psi_n$) and minimizing the quantity (\ref{approx_g}).
A set of harmonic functions $\psi_n$ can be built as
\begin{eqnarray}
\psi_{2n-1} & = & {\cal R}((x_1+ix_2)^n),\quad  \nonumber \\
\psi_{2n} & = & {\cal I}((x_1+ix_2)^n), \quad \rm{for} \quad n\geq 1.
\end{eqnarray}
Or, alternatively,
\begin{eqnarray}
\psi_{2n-1} & = & {\cal R}({\rm exp}(n(x_1+ix_2)/L)),\quad    \nonumber \\
\psi_{2n}   & = & {\cal I}({\rm exp}(n(x_1+ix_2)/L)), \quad \rm{for} \quad n\in\Z^*,
\end{eqnarray}
where ${\cal R}$ and ${\cal I}$ stand for the real and imaginary parts, and $L$ is a constant such that $2\pi\,L$ is larger than the size of the domain.\\

\subsection{Using a wavelet decomposition}
To solve the weak lensing problem on a bounded domain $\Omega$, another possible solution consists in looking for a decomposition of $\boldsymbol \gamma$ of the form:
\begin{equation}
\label{decomp2}
\boldsymbol \gamma= \nabla (p+p_0) +\nabla \times (\omega+\omega_0)=\nabla p' +\nabla \times \omega' .
\end{equation}
By considering a zero B modes constraint as in the previous section,
\begin{equation}
\label{Bconstraintl}
\left[ \begin{array}{c} \kappa_E \\ \kappa_B \end{array} \right] = -\nabla p'+\nabla \times \omega',\quad{\rm with}~
\kappa_B=0,
\end{equation}
the problem can be solved by minimizing the following quantity:
\begin{equation}
\label{approx2}
\min_{p', \omega'} \left\|\boldsymbol \gamma-\nabla p' -\nabla \times \omega' \right\|^2 + \left\| \kappa_B \right\|^2.
\end{equation}

This minimization problem may involve two bases $\{\Psi_{jk}^{\textrm{curl}}\}$ and $\{\Psi_{jk}^{\textrm{div}}\}$ of the divergence-free and curl-free functions on $\Omega$ without any boundary conditions:
\begin{equation}
\label{exp_wav}
\nabla \times (\omega+ \omega_0)=\sum_{jk} d^{\rm curl}_{jk}\Psi_{jk}^{\rm curl}, \quad \nabla (p+p_0)=\sum_{jk} d^{\rm div}_{jk}\Psi_{jk}^{\rm div}.
\end{equation}

This is the option that has been considered in the paper. The basis functions $\Psi_{jk}^{\rm curl}$ and $\Psi_{jk}^{\rm div}$ will consist of wavelets. The construction of these wavelets will be presented in the next section and in Appendix A.
Only a few of the boundary elements $\Psi_{jk}^{\rm curl}$ and $\Psi_{jk}^{\rm div}$ do not satisfy the conditions ${\boldsymbol\gamma}_{\rm div\,0}\cdot{\bf n}=0$ and ${\boldsymbol\gamma}_{\rm curl\,0}\cdot{\boldsymbol\tau}=0$ on the border $\partial\Omega$ as in the decomposition (\ref{divcurlpart}).
Then it is possible to restrain the zero B mode constraint to these border wavelets,  and we call this the B-zero border constraint.

\section{Construction of wavelets for Helmholtz decomposition}
\label{WHD}

In this section, we first review briefly the construction of divergence-free and curl-free wavelets on an unbounded domain. In  Appendix A, we present in detail the bases that we have selected. Then, we exhibit new divergence-free and curl-free wavelets on the square that satisfy suitable boundary conditions.

\subsection{Unbounded domain}

\subsubsection*{Divergence-free and curl-free wavelets}

The compactly supported divergence-free wavelets in $(L^2(\R^n))^n$ originally developed by \citet{wlens:lemarie92} are based on the existence of two different 1D multiresolution analyses (MRA) of $L^2(\R)$ related by differentiation and integration.
Given a 1D  multiresolution analysis $(\varphi_1,\psi_1)$ with $\varphi_1\in C^{1+\epsilon}$ for $\epsilon > 0$, there is another 1D MRA $(\varphi_0,\psi_0)$ such that
\begin{equation}
\label{1Dwav}
\varphi_1'(x)=\varphi_0(x)-\varphi_0(x-1) 
,~~~~~
\psi_1'(x)=4\,\psi_0(x).
\end{equation}

\begin{figure*}[htb]
\centerline{
\includegraphics[width=7cm,height=3.5cm,angle=0]{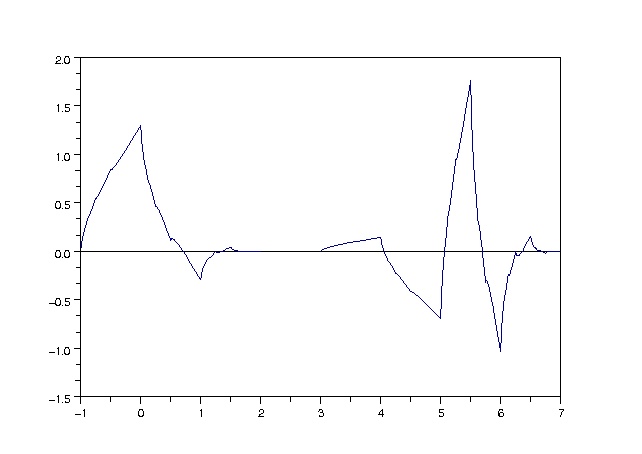}
\includegraphics[width=7cm,height=3.5cm,angle=0]{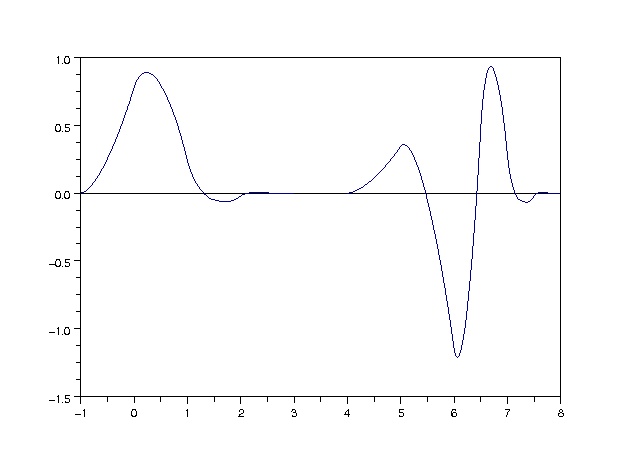}
}
\begin{center}
\begin{tabular}{ccccccc}
$\varphi_0\qquad \qquad \qquad \qquad \qquad \qquad \psi_0$& $ \rm{} $ & $ \rm{} $ & $ \rm{} $ & $ \rm{} $ & $ \rm{} $ & $\varphi_1\qquad \qquad \qquad \qquad \qquad \qquad \psi_1$
\end{tabular}
\end{center}
\caption{Example of two1D MRA related by differentiation and integration. Scaling functions and associated wavelets for biorthogonal MRA with 2 (left) and 3 (right) zero moments.}
\label{mra}
\end{figure*}

\begin{figure*}[htb]
\begin{center}
\begin{tabular}{rccl}
\raisebox{8ex}{$\Phi^{{\rm div}}$} &
\includegraphics[width=4cm,height=4cm]{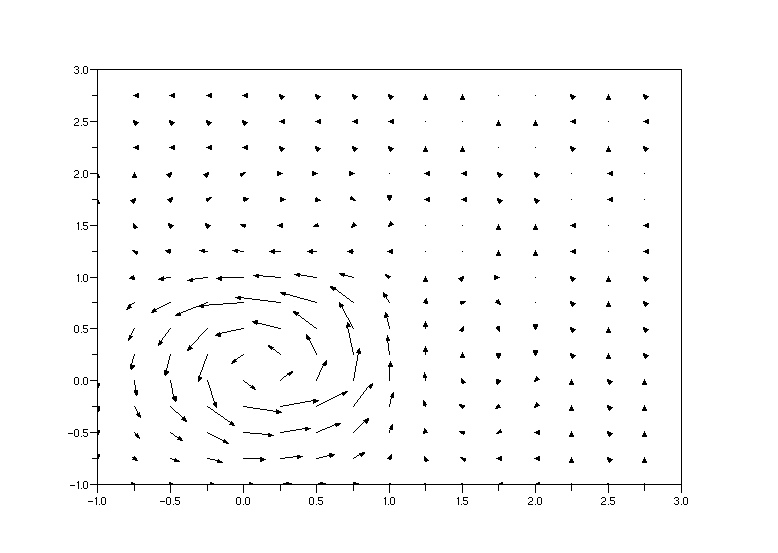}
&
~~~~~~~~\includegraphics[width=4cm,height=4cm]{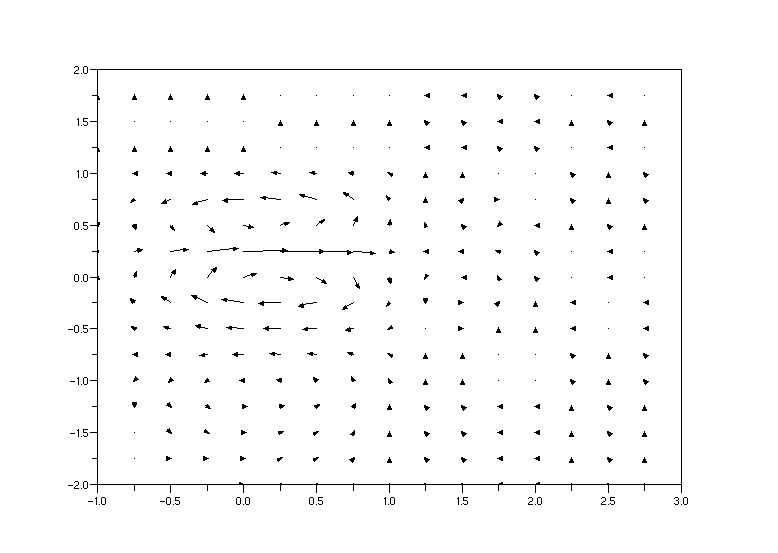}
\raisebox{8ex}{$\Psi^{{\rm div}~(1,0)}$}
\\
\raisebox{8ex}{$\Psi^{{\rm div}~(0,1)}$} &
\includegraphics[width=4cm,height=4cm]{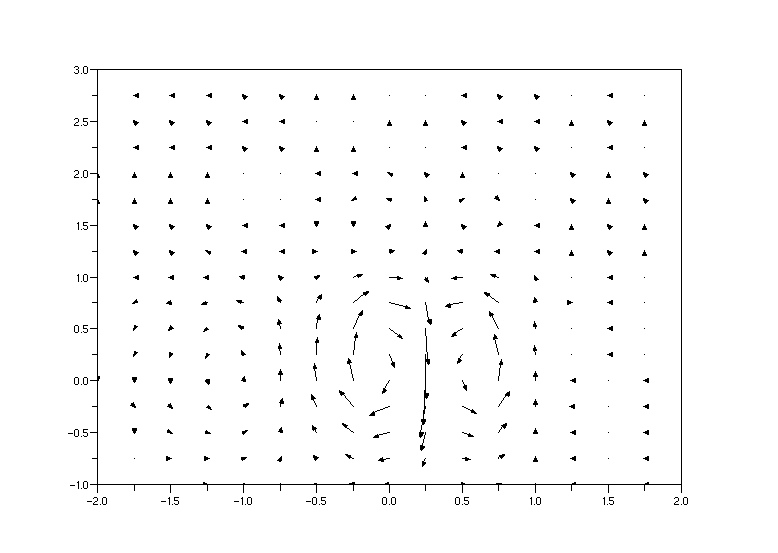}
&
~~~~~~~~\includegraphics[width=4cm,height=4cm]{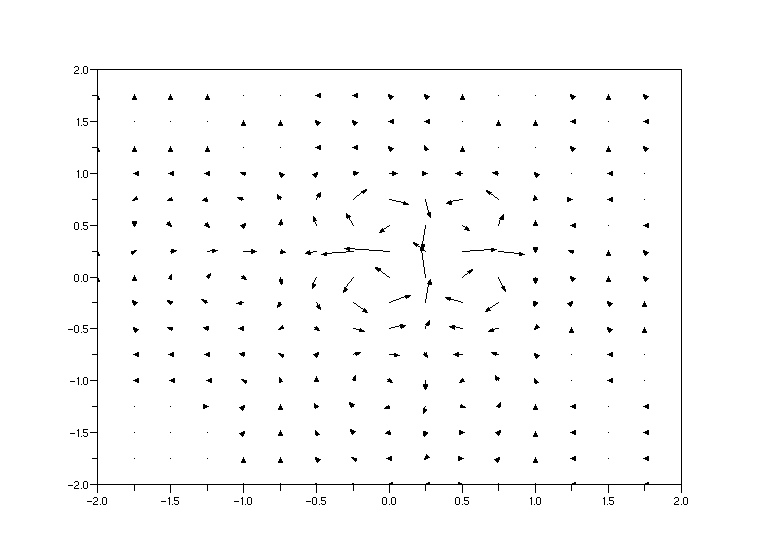}
\raisebox{8ex}{$\Psi^{{\rm div}~(1,1)}$}
\end{tabular}
\caption{Example of a 2D divergence-free wavelet basis designed from the Lemari\'e \& Rieusset (1992) scheme.}
\label{divergence}
\end{center}
\end{figure*}

Different {MRA divergence-free wavelets} can be constructed from different families of 1D MRA's. The construction of MRA curl-free wavelets follows the same approach. Fig.~\ref{mra} shows an example of a family of 1D MRA functions and Fig.~\ref{divergence} shows the corresponding 2D divergence-free wavelets.
These wavelets form a basis of the space of the divergence-free functions on $\R^2$. Nevertheless, they are not suited to implementing the Helmholtz decomposition. 

For this purpose, we prefer to use the {hyperbolic divergence-free and curl-free wavelets} proposed by \citet{wlens:deriaz09}, which form better conditioned bases.
They are based on multidimensional {hyperbolic wavelets} introduced by \citet{wlens:devore98}.
\begin{figure*}[htb]
\begin{center}
\begin{tabular}{rccl}
\raisebox{-16ex}{$\Psi^{\rm div}$} &
\includegraphics[width=4cm,height=4cm,angle=-90]{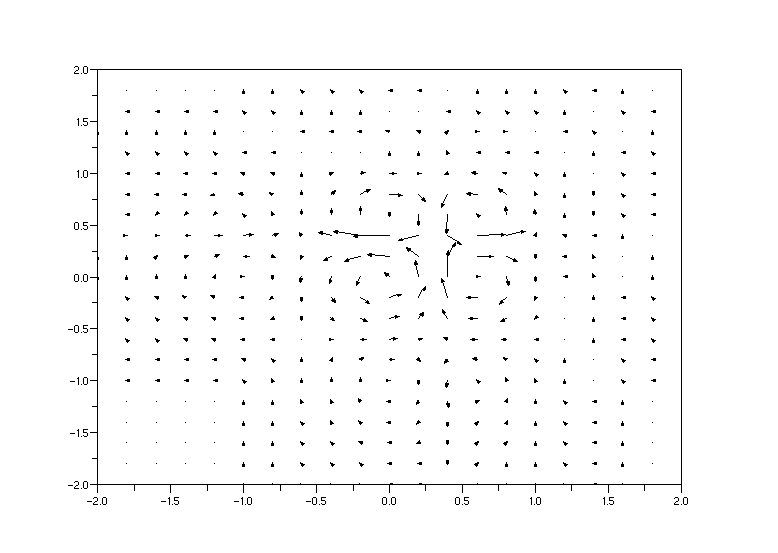}
&
\includegraphics[width=4cm,height=4cm,angle=-90]{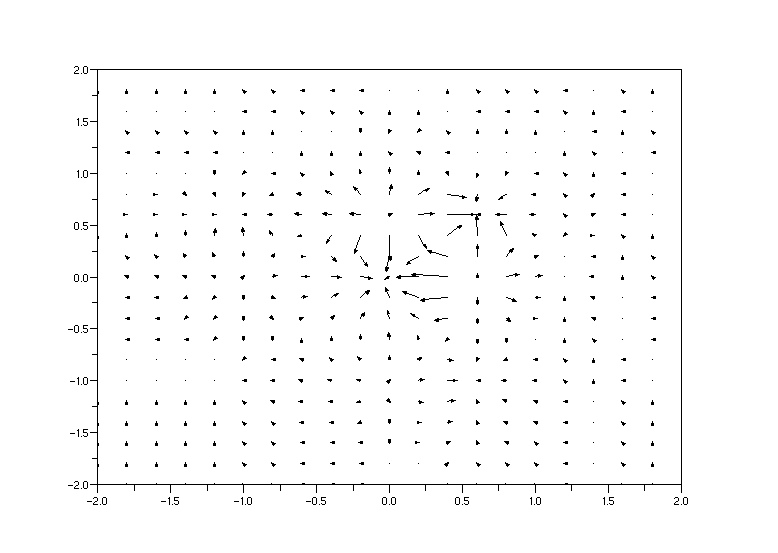} &
\raisebox{-16ex}{$\Psi^{\rm curl}$}
\end{tabular}
\caption{Examples of hyperbolic 2D divergence-free and curl-free wavelets.}
\label{hypdfwfig}
\end{center}
\end{figure*}
The advantage of these wavelets is that there is a fast algorithm to compute the Helmholtz decomposition in terms of divergence-free and curl-free wavelets.


The Helmholtz decomposition in the wavelet domain uses both the divergence-free and curl-free wavelets.
As the divergence-free and curl-free wavelets form biorthogonal bases in their respective spaces, the coefficients $d_{{\bf j,k}}^{\mathrm{curl}}$ and $d_{\bf j,k}^{\textrm{div}}$ corresponding to the divergence-free and curl-free parts (eq. (\ref{exp_wav})) are, in practice, computed with an iterative algorithm \citep[see][ for more details]{wlens:deriaz06}.\\

\subsection{Bounded domain}
\label{Emrecons}
The Helmholtz decomposition (\ref{exp_wav}) may be generalized to bounded domains with non-periodic boundary conditions by the construction of divergence-free and curl-free wavelets on the interval. Since divergence-free and curl-free wavelets are constructed from standard, compactly supported, biorthogonal wavelet bases, they can incorporate boundary conditions. The following section introduces a new construction of wavelets well suited to the Helmholtz decomposition on bounded domains.

\subsubsection{Wavelet transforms on the interval}
\label{wav_interv}
We obtain wavelets on the square by taking tensor products of 1D wavelets on the interval.
The wavelets on the interval we introduce here differ from the original constructions \citep{wlens:meyer90,wlens:cohen93} and rely on the lifting scheme \citep{wlens:daubechies98}.


The properties of these MRA on the interval allow us to apply the wavelet Helmholtz algorithm efficiently on the square: the partition of the unity property facilitates the approximation of the data, and that only $\varphi^{(1)}$ and $\varphi^{(N)}$ differ from $0$ at the boundaries permits us to impose boundary conditions on the wavelets.

\begin{figure*}[htb]
\centerline{
\includegraphics[width=7.6cm,height=6.5cm]{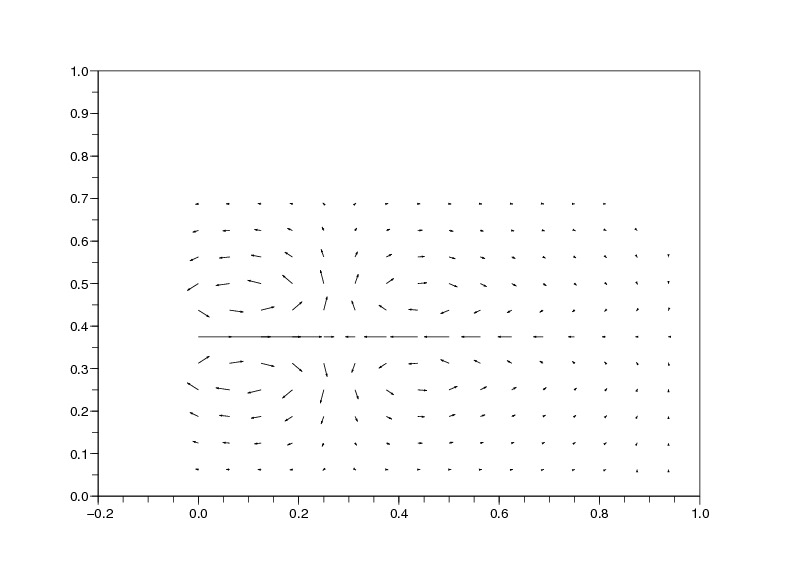}
\includegraphics[width=7.6cm,height=6.5cm]{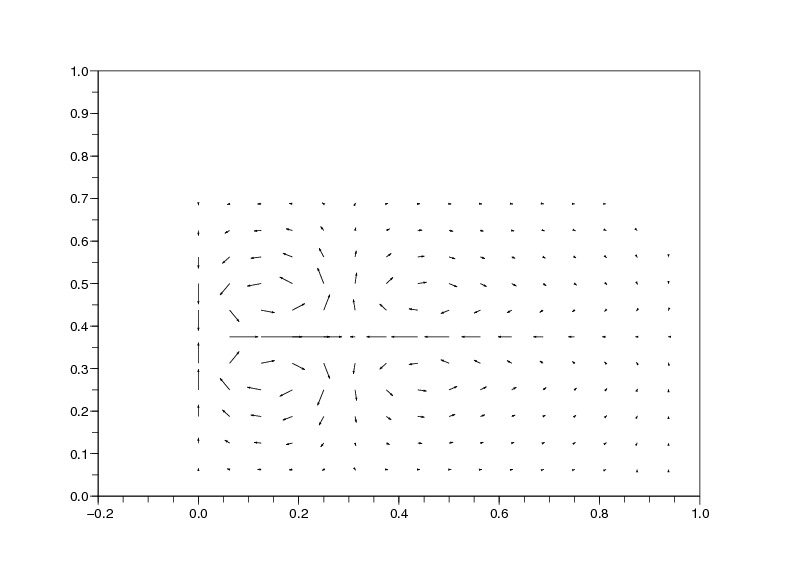}
}
\caption{Example of divergence-free wavelets with a free (left) and a sliding boundary condition (right).}
\label{testpsib.jpg}
\end{figure*}

An example of spline divergence-free wavelets issued from this construction is shown in Fig.~\ref{testpsib.jpg} for free boundary (left) and for sliding conditions (right).
The boundary curl-free wavelets can be constructed similarly.
We applied the resulting boundary divergence-free and curl-free wavelets to perform a Helmholtz decomposition on the bounded domain. In the next section, we present the results obtained with the algorithms that are derived from this construction and presented in Appendix B.

\section{Numerical experiment for weak lensing}
\label{numex}

\subsection{Simulated data}
We have run realistic simulated convergence mass maps derived from N-body cosmological simulations using the RAMSES code \citep{code:teyssier02}. The cosmological model is chosen to be in concordance with the $\Lambda$CDM model. We chose a model with the following parameters close to WMAP : $\Omega_m = 0.3$, $\sigma_8 = 0.9$, $\Omega_L = 0.7$, $h = 0.7$. Each simulation has $256^3$ particles with a box size of 162 Mpc/h. The resulting convergence map covers a field of $2^{\circ}$ x $2^{\circ}$ with 350 x 350 pixels and assumes a galaxy redshift of 1. The overdensities correspond to the halos of groups and clusters of galaxies.

\subsection{Example with no B modes}
\begin{figure*}[htp]
\centerline{
\includegraphics[width=12.5cm,height=6cm]{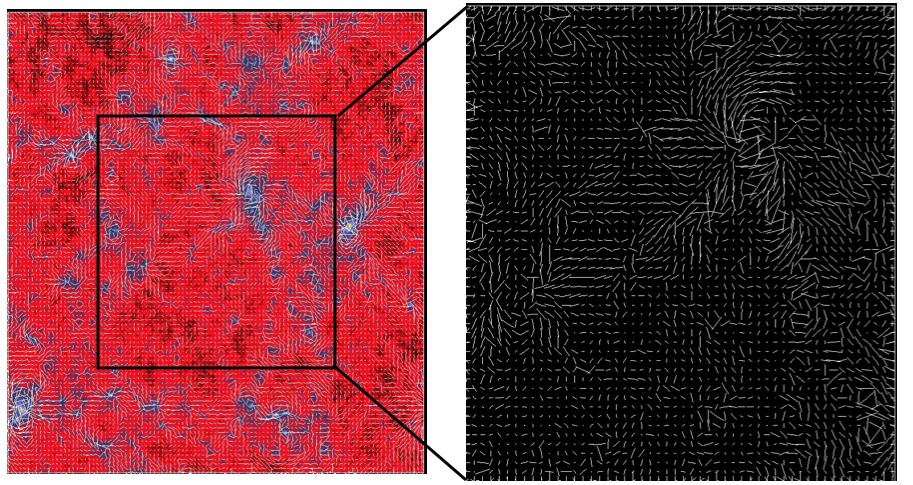}
}
\caption{Left, simulated $\kappa^{S}$ image and its corresponding shear field $\gamma^{S}$. Right, extracted subfield $\gamma^{D}$. The extraction breaks the periodicity of the data.}
\label{fig_kappa_subgamma}
\end{figure*}

In this experiment, we computed the shear map $\gamma^{S}$ from the simulated map $\kappa^{S}$, and we have extracted a sub area $\gamma^{D}$ from $\gamma^{S}$. The lefthand side of Fig.~\ref{fig_kappa_subgamma} shows the simulated $\kappa^{S}$ mass map and its corresponding shear field. The righthand side of Fig.~\ref{fig_kappa_subgamma} shows the subfield $\gamma^{D}$ which is used for the reconstruction.
Then we reconstructed ${\tilde \kappa}_E$ and ${\tilde \kappa}_B$ from $\gamma^{D}$, and compared ${\tilde \kappa}_E$ to $\kappa^{T}$, where $\kappa^{T}$ is the corresponding original area of $\kappa^{S}$. This allows us to study how well the reconstructing methods perform with the border problem.

\begin{figure*}[htp]
\center{
\begin{tabular}{cc}
\includegraphics[width=4.7cm,height=4cm]{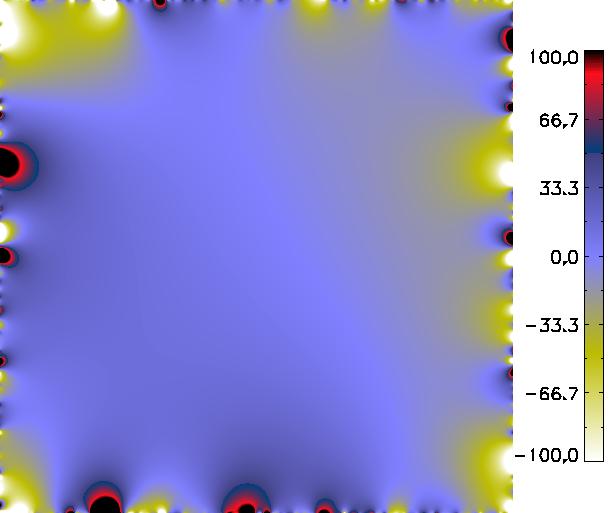} &
\includegraphics[width=4.7cm,height=4cm]{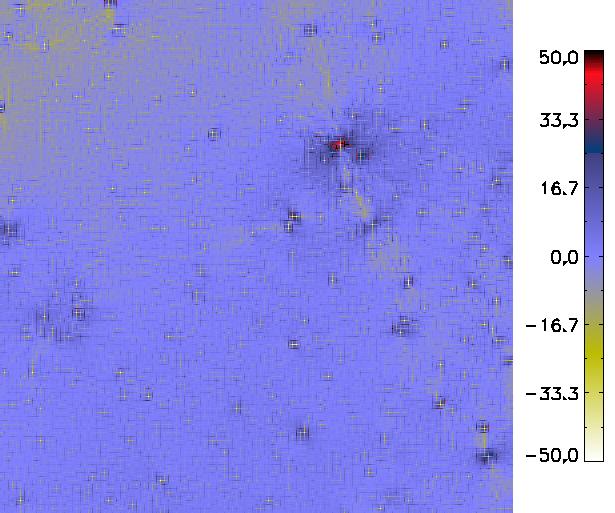} \\
with Fourier & Seitz \& Schneider \\
$\frac{\| {\tilde \kappa}_{E} - \kappa^T \|_{L^2}}{\|\kappa^T\|_{L^2}}=28.0\%$ & 
$\frac{\| {\tilde \kappa}_{E} - \kappa^T \|_{L^2}}{\|\kappa^T\|_{L^2}}=5.27\%$ \\
 & \\
\includegraphics[width=4.7cm,height=4cm]{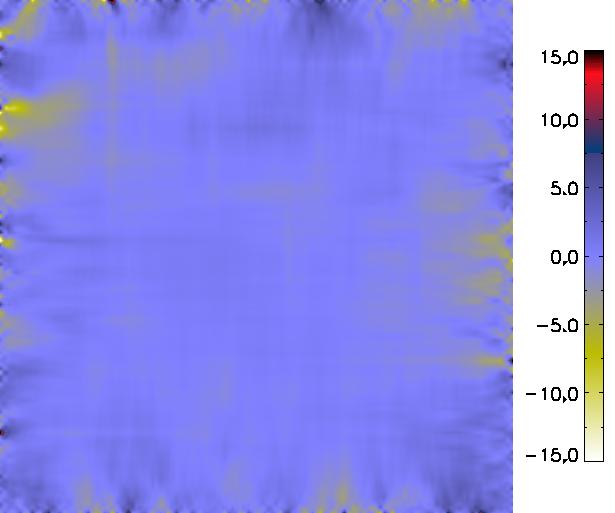} & 
\includegraphics[width=4.7cm,height=4cm]{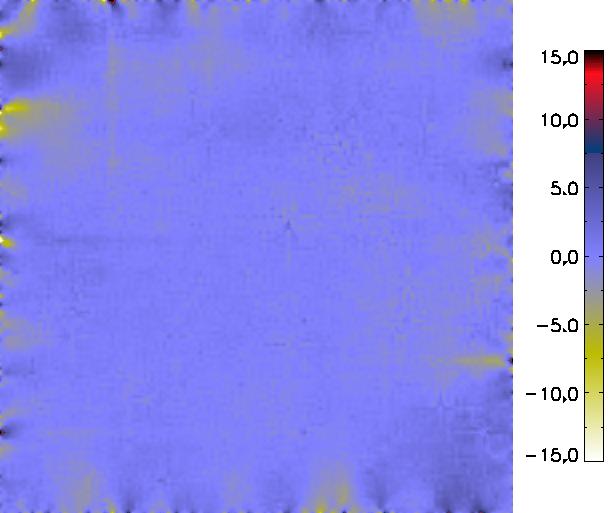} \\
Helmholtz (Zero-B Border Constraint) & Helmholtz (Zero-B Constraint) \\
$\frac{\| {\tilde \kappa}_{E} - \kappa^T \|_{L^2}}{\|\kappa^T\|_{L^2}}=1.2\%$ & $\frac{\| {\tilde \kappa}_{E} - \kappa^T \|_{L^2}}{\|\kappa^T\|_{L^2}}=1.0\%$
\end{tabular}
}
\caption{Reconstruction error maps for four different methods: with the Fourier transform (top left), Seitz \& Schneider (top right), with the constrained Helmholtz decomposition with the constraint only on the boundary (bottom left), and on all the domain (bottom right).}
\label{fig_err_kappa4}
\end{figure*}

Figure~\ref{fig_err_kappa4} shows the error maps relative to four different reconstruction methods: i) the standard FFT approach, ii) Seitz \& Schneider method, iii) the wavelet Helmholtz (50 iterations) with a zero B-mode border constraint, and iv) the wavelet Helmholtz (50 iterations) with a zero B-mode constraint. The error maps were obtained by 
$\varepsilon=  {\tilde \kappa}_{\rm rec}- \kappa^{T}  $.

We calculated the relative percentage error on both $E$ and $B$ modes: 
 $ \epsilon_E = \frac{ \parallel {\tilde \kappa}_E - \kappa^{T} \parallel } { \parallel \kappa^{T} \parallel }$ 
and $ \epsilon_B = \frac{ \parallel {\tilde \kappa}_B \parallel } { \parallel \kappa^{T} \parallel } $, 
and obtain $\epsilon_E =28$ \%, $\epsilon_E =5.3$ \%, $\epsilon_E =1.2$ \% and $\epsilon_E =1.0$ \% for the four methods respectively.
For the B mode component, we obtained $\epsilon_B =30.2$ \% for FFT approach and $\epsilon_B =1.3 $ \% for the wavelet Helmholtz with a zero B modes border constraint.
\begin{figure*}[htp]
\center{
\begin{tabular}{cc}
\includegraphics[width=4.7cm,height=4cm]{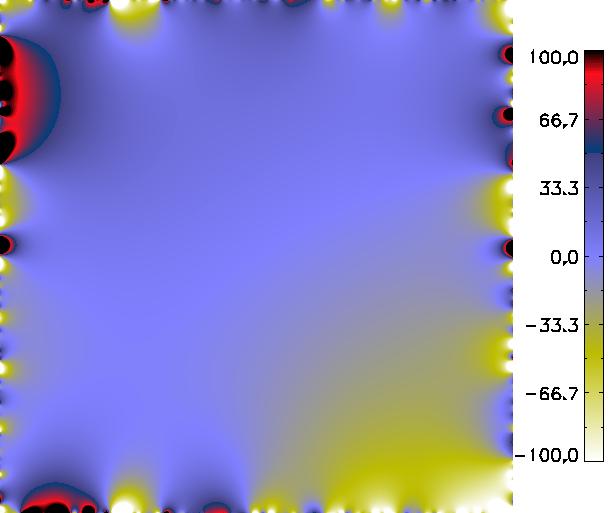} & \includegraphics[width=4.7cm,height=4cm]{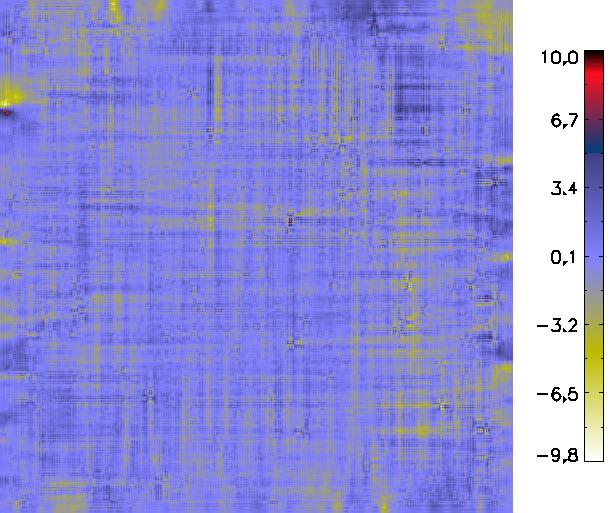} \\
with Fourier & Helmholtz (zero-B border constraint) \\
$\frac{\| {\tilde B}_{rec} \|_{L^2}}{\|\kappa\|_{L^2}}=30.2\%$ & $\frac{\| {\tilde B}_{rec} \|_{L^2}}{\|\kappa\|_{L^2}}=1.3\%$ 
\end{tabular}
}
\caption{B-mode reconstructed maps.}
\label{fig_err_Bmode_rec}
\end{figure*}
Figure~\ref{fig_err_Bmode_rec} shows the reconstructed B-mode map for the Fourier method and the wavelet Helmholtz with a zero B modes border constraint.

This error is not due to the discontinuity on the boundary because we tested the mirroring technique, which consists in symmetrizing the data along the $(Ox)$ and $(Oy)$ axes in order to recover the continuity in exchange to a doubling of the size of the domain.
It produced almost the same error in terms of norm as a straightforward Fourier transform.

From this experiment, we can conclude that the wavelet Helmholtz method allows us to dramatically reduce the error relative to the border effect, when no B modes contaminate our data.

\subsection{Example with B modes}
In this experiment, we have added B modes. The B-mode map consists in a Gaussian with a maximum equal to 10\% of the maximum of $\kappa^T$ from the mass map.
Since the constrained wavelet Helmholtz decomposition assumes no B modes, or zero-border B modes, it is interesting to see what happens when we introduce some actual B modes into the model.

\begin{figure*}[htb]
\begin{tabular}{ccc}
\includegraphics[width=4.7cm,height=4cm]{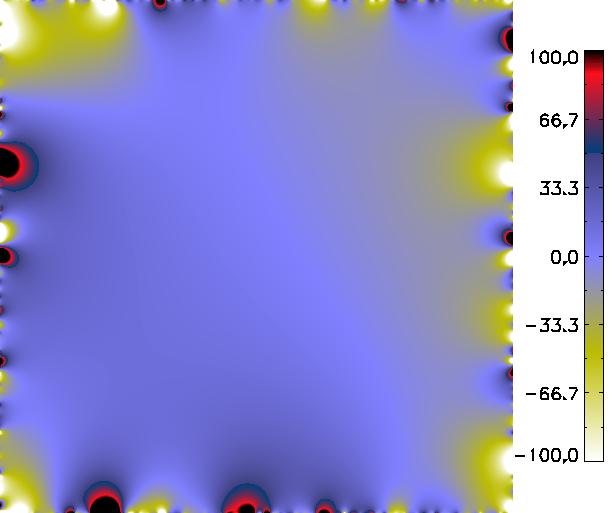} &
\includegraphics[width=4.7cm,height=4cm]{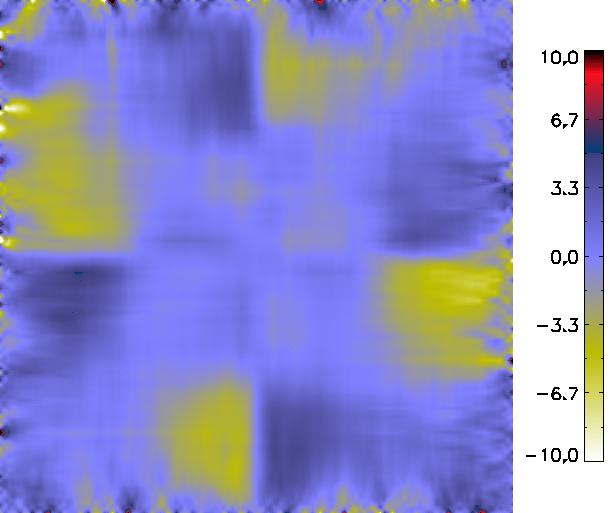} &
\includegraphics[width=4.7cm,height=4cm]{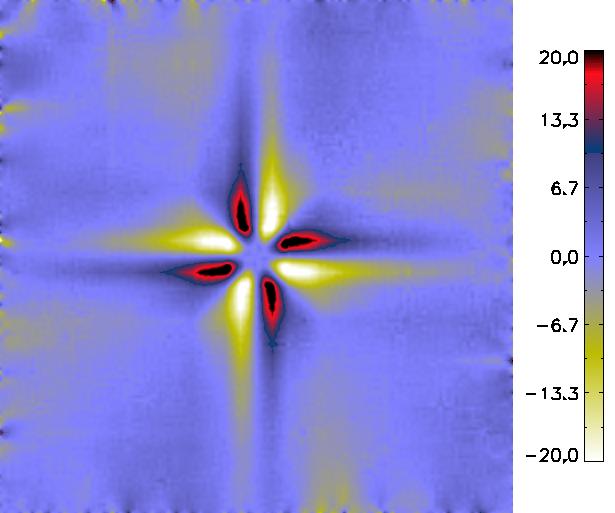}\\
with Fourier & Helmholtz (B-zero border constraint) & Helmholtz (B-zero constraint) \\
$\frac{\| {\tilde \kappa}_{E} - \kappa^T \|_{L^2}}{\|\kappa^T\|_{L^2}}=28.0\%$ & $\frac{\| {\tilde \kappa}_{E} - \kappa^T \|_{L^2}}{\|\kappa^T\|_{L^2}}=2.1\%$ & $\frac{\| {\tilde \kappa}_{E} - \kappa^T \|_{L^2}}{\|\kappa^T\|_{L^2}}=4.0\%$
\end{tabular}
\caption{Maps of the reconstruction error on $\kappa_{E}$ for three different methods when B modes contaminate the data.}
\label{fig_err_kappa_with_B}
\end{figure*}

Figure~\ref{fig_err_kappa_with_B} shows the reconstruction error maps with the three methods. We can see that the reconstruction error when using Fourier is not amplified in the presence of B modes, while the error increases when using the wavelet Helmholtz with a constraint on the B modes.
This is expected since the model is not completely true anymore. We note, however, that even if the error increases, it is still much below the FFT error.
In this case, wavelet Helmholtz with zero border B modes is more robust than wavelet Helmholtz with zero B modes.

\begin{figure*}[htb]
\begin{center}
\begin{tabular}{ccc}
\includegraphics[width=4.7cm,height=4cm]{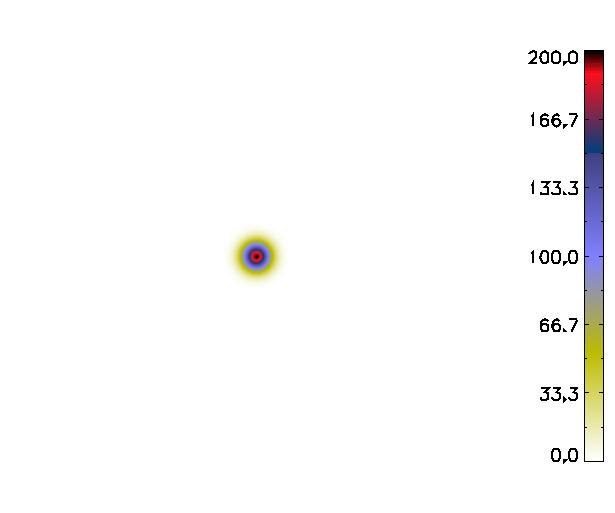} &
\includegraphics[width=4.7cm,height=4cm]{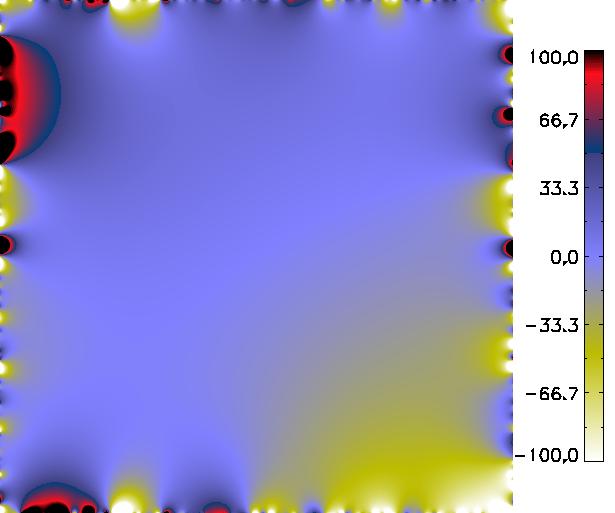} & 
\includegraphics[width=4.7cm,height=4cm]{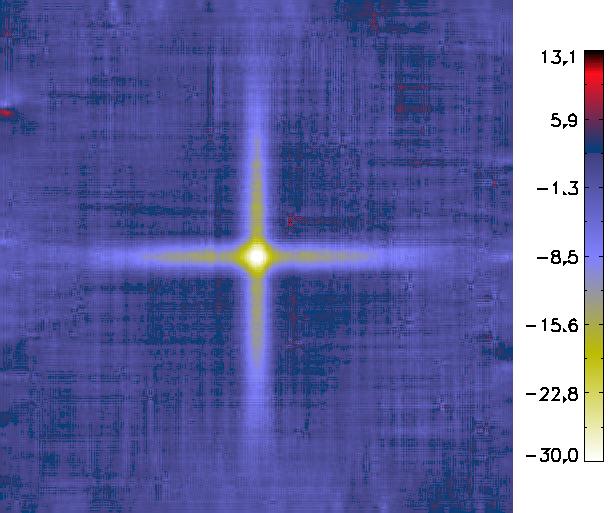} \\
Original B modes map & Error map with Fourier & Error map with Helmholtz (B-Zero Border Constraint) \\
 & $\frac{\| {\tilde \kappa}_{B} - \kappa_B^T \|_{L^2}}{\|\kappa^T\|_{L^2}}=30.2\%$ & $\frac{\| {\tilde \kappa}_{B} - \kappa_B^T \|_{L^2}}{\|\kappa^T\|_{L^2}}=3.2 \%$ 
\end{tabular}
\end{center}
\caption{Original B mode and reconstructed error maps for the B modes map $\kappa_{B}$.}
\label{fig_err_Bmode_rec2}
\end{figure*}
Figure~\ref{fig_err_Bmode_rec2} shows the original B mode map and the reconstructed error maps  for both the Fourier transform and the wavelet Helmholtz decomposition with a zero B-mode border constraint.
From this experiment, we can conclude that the wavelet Helmholtz method with zero-border B-mode constraint 
is the best trade-off when some residual B modes may contaminate the data.
 



\section{Conclusion}
\label{concl}

We have sorted out a direct and original relationship between the Helmholtz decomposition and the E/B mode reconstruction. The Helmholtz decomposition can be performed without the FFT, by using divergence-free and curl-free wavelets.
This wavelet Helmholtz decomposition is iterative, and we have shown that we have different strategies to handle the border, thanks to the versatility of wavelets.
This means that, periodic borders are not a curse anymore.
Indeed, we were able to design two kinds of Helmholtz decompositions, one imposing B modes as equal to 0, and another imposing B modes as zero only on the border.
We have shown that both approaches outperform the FFT-based standard E/B mode reconstruction, which suffers from the serious drawback of imposing periodic border conditions that concentrate in the low frequencies. The
smaller the field, the more important the bias introduced by this periodic border condition. In our experiments, using the Helmholtz wavelet
decomposition reduces dramatically the error. 
{We have shown that the wavelet Helmholtz decomposition outperforms also Seitz and Schneider method \citep{wlens:seitz96}.}
Our approach also offers the advantage of no needing to pixelate the shear map. 
Indeed, both divergence-free and curl-free wavelet transforms can be applied directly on the shear catalog without having to pixelate first on a uniform grid.

However, at this stage, our wavelet Helmholtz cannot be properly used yet on real data. Indeed, two main issues have not been addressed in this paper,
the noise and the missing data. Both may be problematic: stability, convergence, and unicity of the solution must be studied in detail.
This will be done in the future.


\begin{acknowledgements}
The authors thank Peter Schneider for kindly providing his code, and Adrienne Leonard for her help in using it.
This work was supported by the French National Agency for Research (ANR -08-EMER-009-01) and 
the European Research Council grant SparseAstro (ERC-228261). 
\end{acknowledgements}

\section*{Appendix A: Divergence-free wavelets on a square domain}

\subsection*{Expression of the divergence-free and curl-free wavelets}

With the 1D wavelets of Eq. (\ref{1Dwav}), the 2D {\sl isotropic} divergence-free scaling function is defined as \citep{wlens:lemarie92}
\begin{equation}
\Phi^{\textrm{curl}}(x_1,x_2)=\left( \begin{array}{l}%
\varphi_1(x_1)\varphi_1'(x_2) \\
-\varphi_1'(x_1)\varphi_1(x_2)
\end{array}\right),
\end{equation}
and the details on each scale are obtained from the three wavelets\\

- Horizontal wavelet: \\
$\Psi_{j,\k}^{\textrm{curl}\,(1,0)}(x_1,x_2)=\left( \begin{array}{l} \psi_1(2^jx_1-k_1)\varphi_1'(2^jx_2-k_2) \\ -\psi_1'(2^jx_1-k_1)\varphi_1(2^jx_2-k_2) \end{array}\right) $;\\

- Vertical wavelet: \\
$\Psi_{j,\k}^{\textrm{curl}\,(0,1)}(x_1,x_2)=\left( \begin{array}{l} -\varphi_1(2^jx_1-k_1)\psi_1'(2^jx_2-k_2) \\ \varphi_1'(2^jx_1-k_1)\psi_1(2^jx_2-k_2) \end{array}\right) $;\\

- Diagonal wavelet:  \\
$\Psi_{j,\k}^{\textrm{curl}\,(1,1)}(x_1,x_2)=\left( \begin{array}{l} \psi_1(2^jx_1-k_1)\psi_1'(2^jx_2-k_2) \\ -\psi_1'(2^jx_1-k_1)\psi_1(2^jx_2-k_2) \end{array}\right) $.\\


The hyperbolic divergence-free vector wavelets are expressed as
\begin{itemize}
\item Divergence-free wavelets:
\begin{equation}
\Psi_{{\bf j},{\bf k}}^{\textrm{curl}}(x_1,x_2)=\left( \begin{array}{l}
{2^{j_2}}\psi_1({2^{j_1}}x_1-k_1)\psi_0({2^{j_2}}x_2-k_2) \\
-{2^{j_1}}\psi_0({2^{j_1}}x_1-k_1)\psi_1({2^{j_2}}x_2-k_2)\end{array}\right);
\end{equation}
\item Curl-free wavelets:
\begin{equation}
\Psi_{{\bf j},{\bf k}}^{\textrm{div}}(x_1,x_2)=\left( \begin{array}{l}
{2^{j_1}}\psi_0({2^{j_1}}x_1-k_1)\psi_1({2^{j_2}}x_2-k_2) \\
{2^{j_2}}\psi_1({2^{j_1}}x_1-k_1)\psi_0({2^{j_2}}x_2-k_2)\end{array}\right)
\end{equation}
with the scale parameter ${\bf j}=(j_1,j_2)\in\Z^2$ and the position parameter
${\bf k}=(k_1,k_2)\in\Z^2$. The wavelet is localized at $(2^{-j_1}k_1,2^{-j_2}k_2)$.
\end{itemize}


The fast wavelet decomposition in these vector wavelet bases proceeds as follows.
First, in 2D, we need to define two differentiable wavelet bases \citep{wlens:deriaz06}:
\begin{eqnarray}
\begin{array}{lll}
\Psi_{1\,{\bf j},{\bf k}}(x_1,x_2) & = & \left( \begin{array}{l} \psi_1(2^{j_1}x_1-k_1)\psi_0(2^{j_2}x_2-k_2) \\ 0
\end{array} \right)  \\
\tilde{\Psi}_{1\,{\bf j},{\bf k}}(x_1,x_2) & = & \left( \begin{array}{l} \psi_0(2^{j_1}x_1-k_1)\psi_1(2^{j_2}x_2-k_2) \\ 0
\end{array} \right) \\
\Psi_{2\,{\bf j},{\bf k}}(x_1,x_2) & = & \left( \begin{array}{l} 0 \\ \psi_0(2^{j_1}x_1-k_1)\psi_1(2^{j_2}x_2-k_2) \end{array} \right) \\
\tilde{\Psi}_{1\,{\bf j},{\bf k}}(x_1,x_2) & = & \left( \begin{array}{l} 0 \\ \psi_1(2^{j_1}x_1-k_1)\psi_0(2^{j_2}x_2-k_2)
\end{array} \right).
\end{array}
\end{eqnarray}
We decompose the field $\gamma$ in these bases.  
Then, as the divergence-free and curl-free wavelets are defined by \\
$
\Psi_{{\bf j},{\bf k}}^{\textrm{curl}}=2^{j_2}\Psi_{1\,{\bf j},{\bf k}}-2^{j_1}\Psi_{2\,{\bf j},{\bf k}}$,  and \\
$ \Psi_{{\bf j},{\bf k}}^{\textrm{div}}=2^{j_1}\tilde{\Psi}_{1\,{\bf j},{\bf k}}+2^{j_2}\tilde{\Psi}_{2\,{\bf j},{\bf k}}$. \\
The vector wavelet expansion is obtained by a change of basis (cf \citep{wlens:deriaz06}).

One can notice that the divergence-free wavelets correspond to the rotational of the wavelet basis $\{\frac{1}{4}(\psi_{1\,j_1,k_1}\otimes\psi_{1\,j_2,k_2})\}$ and the curl-free wavelets correspond simply to its gradient. 


\subsection*{Multi-Resolution Analyses on the interval linked by derivation}
\label{sec_wav_interv}

\begin{figure*}[htb]
\begin{tabular}{ccc}
\includegraphics[width=4.8cm,height=3.3cm]{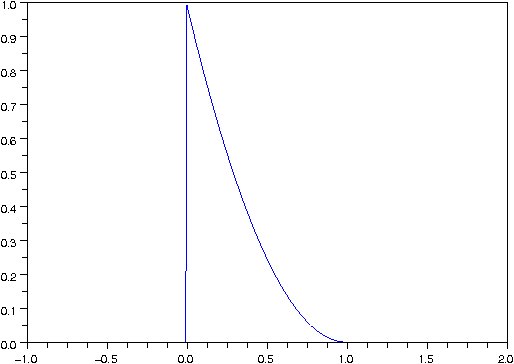}&
\includegraphics[width=4.8cm,height=3.3cm]{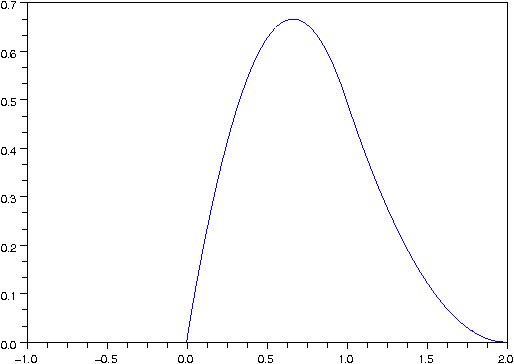}&
\includegraphics[width=4.8cm,height=3.3cm]{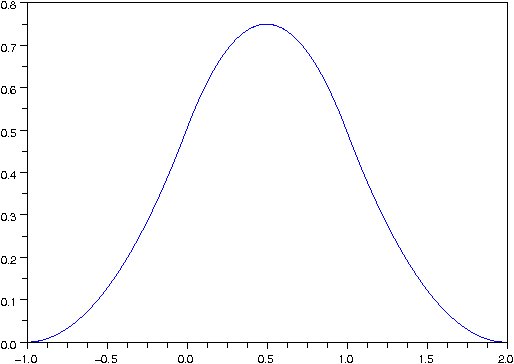} \\
$\varphi_1^{(1)}$ & $\varphi_1^{(2)}$ & $\varphi_1$
\end{tabular}
\caption{\label{phi_1b} Quadratic spline scaling functions adapted to the left boundary of the interval.}
\end{figure*}

We extend this construction to bounded domains using special 1D wavelets on the interval.
To obtain the MRAs on $[0,1]$, we modify the filters of multiresolution analyses on $\R$ on the boundaries.
The associated scaling functions (father wavelets) form a partition of unity in the sense that their sum is equal to one.
For instance the quadratic spline filter,
\begin{equation}
\varphi_1(x)=\frac{1}{4}\varphi_1(2x+1)+\frac{3}{4}\varphi_1(2x)+\frac{3}{4}\varphi_1(2x-1)+\frac{1}{4}\varphi_1(2x-2),
\end{equation}
is modified on the left boundary to
\begin{equation}
\varphi_1^{(1)}(x)=\varphi_1^{(1)}(2x)+\frac{1}{2}\varphi_1^{(2)}(2x),
\end{equation}
and
\begin{equation}
\varphi_1^{(2)}(x)=\frac{1}{2}\varphi_1^{(2)}(2x)+\frac{3}{4}\varphi_1(2x-1)+\frac{1}{4}\varphi_1(2x-2).
\end{equation}
We thus we obtain the corresponding scaling functions plotted in Fig.~\ref{phi_1b}.

To form the divergence-free and curl-free wavelets, we need two MRAs linked by derivation.
In the following, we present a method to integrate an MRA that was partially derived from \citet{wlens:stevenson10}.

We start from a Multi-Resolution Analysis (MRA) $(\varphi_0^{(1)},\varphi_0^{(2)},\dots,\varphi_0^{(M)})$ on the interval $[0,N]$,
verifying
\begin{equation}
\sum_{i=1}^{M}\varphi_0^{(i)}(x)=\mathbf{1}_{[0,N]}(x), \quad {\rm and}\quad \varphi_0^{(1)}(0)=\varphi_0^{(M)}(N)=1.
\end{equation}
We put
\begin{equation}
c_0^{(i)}=\int_0^N\varphi_0^{(i)}(x)\,dx.
\end{equation}
Then we define the MRA $(\varphi_1^{(1)},\varphi_1^{(2)},\dots,\varphi_1^{(M+1)})$ by
\begin{equation}
\varphi_1^{(1)}(x)=1-\frac{1}{c_0^{(1)}}\int_0^x\varphi_0^{(1)}(t)\,dt,
\end{equation}
\begin{equation}
\varphi_1^{(i)}(x)=\int_0^x\left( \frac{1}{c_0^{(i-1)}}\varphi_0^{(i-1)}(t)-\frac{1}{c_0^{(i)}}\varphi_0^{(i)}(t)\right)\,dt,~~~~
\end{equation}
${\rm for}~~ i=2,\dots,M$,

\begin{equation}
\varphi_1^{(M+1)}(x)=\frac{1}{c_0^{(M)}}\int_0^x\varphi_0^{(M)}(t)\,dt.
\end{equation}
Then the set $\{\varphi_1^{(i)}\}$ forms a partition of the unity,
\begin{equation}
\sum_{i=1}^{N+n}\varphi_1^{(i)}(x)=\mathbf{1}_{[0,N]}(x),
\end{equation}
and satisfies $\varphi_1^{(1)}(0)=\varphi_1^{(M+1)}(N)=1$. Thanks to this construction, to any wavelet $\psi_1^{\ell}$ defined in the MRA formed by
$(\varphi_1^{(1)},\dots,\varphi_1^{(M+1)})$ we can associate a wavelet $\psi_0^{\ell}$ in the MRA formed
by $(\varphi_0^{(1)},\dots,\varphi_0^{(M)})$ by the relation ${\psi_1^{\ell}}'=4\psi_0^{\ell}$.

In the example of Section \ref{wav_interv}, the corresponding MRA $(\varphi_0^{(1)},\dots,\varphi_0^{(M)})$ are the
linear splines on the interval. The filter is given by
\begin{equation}
\varphi_0(x)=\frac{1}{2}\varphi_0(2x+1)+\varphi_0(2x)+\frac{1}{2}\varphi_0(2x-1),
\end{equation}
and the left boundary filter by
\begin{equation}
\varphi_0^{(1)}(x)=\varphi_0^{(1)}(2x)+\frac{1}{2}\varphi_0(2x-1),
\end{equation}
\begin{figure*}[htb]
\begin{center}
\begin{tabular}{cc}
\includegraphics[width=4.8cm,height=3.3cm]{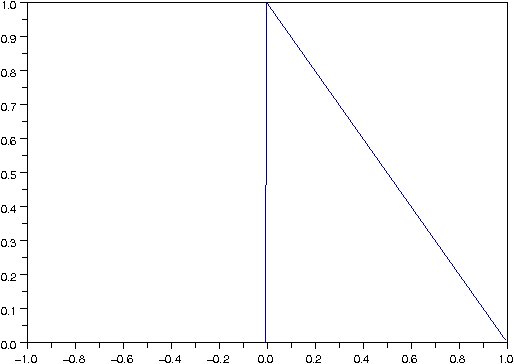}&
\includegraphics[width=4.8cm,height=3.3cm]{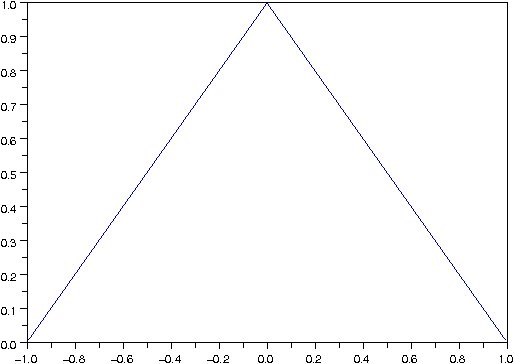} \\
$\varphi_0^{(1)}$ & $\varphi_0$
\end{tabular}
\end{center}
\caption{\label{phi_0b} Linear spline scaling functions adapted to the interval.}
\end{figure*}


\subsection*{Divergence-free wavelets on a bounded domain}
From the differentiation relation introduced above
\begin{equation}
{\psi_1^{\rm b}}'=4\,\psi_0^{\rm b},
\end{equation}
we form the boundary divergence-free wavelets:
\begin{equation}
\label{psi^b}
\Psi_{{\bf j},(0,k_2)}^{\textrm{curl}\,\rm b}(x_1,x_2)=\left( \begin{array}{l}
{2^{j_2}}\psi_1^{\rm b}({2^{j_1}}x_1)\psi_0({2^{j_2}}x_2-k_2) \\
-{2^{j_1}}\psi_0^{\rm b}({2^{j_1}}x_1)\psi_1({2^{j_2}}x_2-k_2)\end{array}\right).
\end{equation}
For these boundary divergence-free wavelets, the border conditions are given by $\psi_1^{\rm b}(0)$, perpendicular to the border and by $\psi_0^{\rm b}(0)$, tangential to the border.

\section*{Appendix B: Algorithms}

In this appendix, we present a new algorithm based on the wavelet Helmholtz decomposition that improves the weak lensing E/B mode decomposition.

\subsection*{Wavelet Helmholtz decomposition algorithm}
\label{wav_Helm_alg}

The Helmholtz decomposition may be viewed as the following orthogonal space splitting:
\begin{equation}
(L^2(\R^2))^2={\textbf{H}}_{\mathrm{div}\,0}(\R^2)\oplus^{\bot} {\textbf{H}}_{\textrm{curl}\,0}(\R^2),
\end{equation}
where ${\textbf{H}}_{\mathrm{div}\,0}(\R^2)$ is the space of divergence-free vector functions and ${\textbf{H}}_{\textrm{curl}\,0}(\R^2)$ is the space of curl-free vector functions.
According to the wavelet decomposition \citep{wlens:deriaz06},
we have the following direct sums of MRA spaces:
\begin{equation}
\begin{array}{c} (L^2(\R^2))^2={\textbf{H}}_{\mathrm{div}\,0}(\R^2) \oplus \textbf{H}^\mathrm{c}_\mathrm{div}(\R^2), \\
(L^2(\R^2))^2=\textbf{H}^\mathrm{c}_\mathrm{curl}(\R^2) \oplus {\textbf{H}}_{\mathrm{curl}\,0}(\R^2). \end{array}
\end{equation}
{These correspond to the decompositions in the divergence-free and curl-free wavelet bases, and, denoting $\mathrm{P}_{\mathrm{div}\,0}$ as the non orthogonal wavelet projector on ${\textbf{H}}_{\mathrm{div}\,0}$ --see \citep{wlens:deriaz06} for its definition-- we have}
\begin{equation}
\label{decompnonorth}
{\boldsymbol\gamma}=\mathrm{P}_{\mathrm{div}\,0}\,{\boldsymbol\gamma}+\mathrm{P}_{\mathrm{div}}^\mathrm{c}\,{\boldsymbol\gamma},~~~~~
{\boldsymbol\gamma}=\mathrm{Q}_{\mathrm{curl}}^\mathrm{c}\,{\boldsymbol\gamma}+\mathrm{Q}_{\mathrm{curl}\,0}\,{\boldsymbol\gamma},
\end{equation}
where $\mathrm{P}^\mathrm{c}_{\mathrm{div}}$ and $\mathrm{Q}^\mathrm{c}_{\mathrm{curl}}$ are complementary projectors of $\mathrm{P}_{\mathrm{div}\,0}$ and $\mathrm{Q}_{\mathrm{curl}\,0}$, respectively. 

Unfortunately, the projectors on the divergence-free wavelet and curl-free wavelet bases are {\it biorthogonal} projectors, so we need an iterative method to provide ${\boldsymbol\gamma}_{{\rm div}\,0}$ and ${\boldsymbol\gamma}_{{\rm curl}\,0}$ and to write
\begin{equation}
\label{gamma_dec_d_c}
{\boldsymbol\gamma}={\boldsymbol\gamma}_{\mathrm{div}\,0}+{\boldsymbol\gamma}_{\mathrm{curl}\,0}.
\end{equation}
Since the wavelet bases are conditioned well, it is possible to apply them recursively in order to construct sequences ${\boldsymbol\gamma}_{{\rm div}\, 0}^p$ and ${\boldsymbol\gamma}_{{\rm curl}\, 0}^p$, which converge to ${\boldsymbol\gamma}_{\mathrm{div}\,0}$ and ${\boldsymbol\gamma}_{\mathrm{curl}\, 0}$ in a straight way. 
Then, in order to decompose the data $\boldsymbol\gamma$ into its divergence-free and curl-free parts $\boldsymbol\gamma=\underbrace{\P\boldsymbol\gamma}_{\u}+\underbrace{\Q\boldsymbol\gamma}_{\v}$, we have implemented the following iterative algorithm:

\begin{itemize}
\item[1-] Set $\u^0$ = $\v^0$ = 0 and the maximum number of iterations to $I_{max}$.
\item[2-] Set i = 0, while $|| \boldsymbol\gamma - \u^i - \v^i ||^2 > \epsilon$ and $i < I_{max}$ then iterate:
\begin{itemize}
\item[1-] Set $\boldsymbol\gamma_{\rm{res}}^i = \boldsymbol\gamma - \u^i - \v^i $
\item[2-] Set $\u^{i+1} = \u^{i}+\mathrm{P}_{\mathrm{div}\,0}\,\boldsymbol\gamma^i_{\rm{res}}$
\item[3-] Set $\boldsymbol\gamma_{\rm{res}}^i = \boldsymbol\gamma - \u^{i+1} - \v^i $
\item[4-] Set $\v^{i+1} = \v^{i}+\mathrm{Q}_{\mathrm{curl}\,0}\,\boldsymbol\gamma^i_{\rm{res}}$
\item[5-] Set $i = i+1$
\end{itemize}
\end{itemize}

The algorithm converges when the residual $\boldsymbol\gamma_{\rm{res}}^i$ goes to $0$ and has been proven numerically.
We tested this algorithm with periodic boundary conditions and obtained almost the same result as with the Fourier transform. Another point is that this algorithm still converges on a bounded domain with free boundary condition wavelets (see the left panel of fig.~\ref{gamb.jpg}).

\subsection*{Weak lensing reconstruction on a bounded domain}

In this section, we present the iterative algorithm that aims at adding the constraint $\kappa_{B}=0$ in the wavelet Helmholtz decomposition algorithm to improve the reconstruction of the E modes convergence $\kappa_{E}$ in a bounded domain:
\begin{equation}
\boldsymbol\gamma=\P\boldsymbol\gamma+\Q\boldsymbol\gamma, \quad\quad
\left(\begin{array}{c} \kappa_{E} \\ 0 \end{array}\right)=-\P\boldsymbol\gamma+\Q\boldsymbol\gamma.
\end{equation}
The algorithm that we have implemented is
\begin{itemize}
\item[1-] Set $\u^0$ = $\v^0$ = 0 and the maximum number of iterations $I_{max}$.
\item[2-] Set i = 0, while $|| \boldsymbol\gamma - \u^i - \v^i ||^2 > \epsilon$ and $i < I_{max}$ then iterate:
\begin{itemize}
\item[1-] Set $\boldsymbol\gamma_{\rm{res}}^i = \boldsymbol\gamma - \u^i - \v^i $
\item[2-] Set $\u^{i+\frac{1}{2}} = \u^{i}+\mathrm{P}_{\mathrm{div}\,0}\,\boldsymbol\gamma^i_{\rm{res}}$
\item[3-] Set $ \w^{i} = \left(\begin{array}{c} 0 \\ -u_2^{i+\frac{1}{2}} \end{array}\right)+\left(\begin{array}{c} 0 \\ v_2^i \end{array}\right)$
\item[4-] Set $\u^{i+1} = \u^{i+\frac{1}{2}}+\mathrm{P}_{\mathrm{div}\,0}\,\w^{i}$ 
\item[5-] Set $\boldsymbol\gamma_{\rm{res}}^{i+\frac{1}{2}} = \boldsymbol\gamma - \u^{i+1} - \v^i $
\item[6-] Set $\v^{i+\frac{1}{2}} = \v^{i}+\mathrm{Q}_{\mathrm{curl}\,0}\,\boldsymbol\gamma_{\rm{res}}^{i+\frac{1}{2}}$
\item[7-] Set $ \w^{i+\frac{1}{2}} = \left(\begin{array}{c} 0 \\ -u_2^{i+1} \end{array}\right)+\left(\begin{array}{c} 0 \\ v_2^{i+\frac{1}{2}} \end{array}\right)$
\item[8-] Set $\v^{i+1} = \v^{i+\frac{1}{2}}-\mathrm{Q}_{\mathrm{curl}\,0}\,\w^{i+\frac{1}{2}}$
\item[9-] Set $i = i+1$
\end{itemize}
\end{itemize}

This algorithm converges with a slow decay rate.\\

An alternative to this algorithm consists in restricting the B modes constraint only to the boundary wavelets $\{\Psi_{{\bf j},k}^{\rm b}\}$. 
With such an approach, we are able to isolate some of the B modes that are not harmonic.
The alternative algorithm will be 
\begin{itemize}
\item[1-] Set $\u^0$ = $\v^0$ = 0 and the maximum number of iterations to $I_{max}$.
\item[2-] Set i = 0, while $|| \boldsymbol\gamma - \u^i - \v^i ||^2 > \epsilon$ and $i < I_{max}$ then iterate:
\begin{itemize}
\item[1-] Set $\boldsymbol\gamma_{\rm{res}}^i = \boldsymbol\gamma - \u^i - \v^i $
\item[2-] Set $\u^{i+\frac{1}{2}} = \u^{i}+\mathrm{P}_{\mathrm{div}\,0}\,\boldsymbol\gamma^i_{\rm{res}}$ with: $\u^{i+\frac{1}{2}}=\sum_{jk}u_{jk}^{i+\frac{1}{2}}\Psi_{jk}^{\textrm{div}}$
\item[3-] Compute $\u_{\delta \Omega}^{i+\frac{1}{2}}=\sum_{jk~{\rm s.t.}~\Psi_{jk\,|\partial\Omega}^{\textrm{div}}\neq 0}u_{jk}^{i+\frac{1}{2}}\Psi_{jk}^{\textrm{div}}$
\item[4-] Set $ \w^{i} = \left(\begin{array}{c} 0 \\ -u_{\delta\Omega~2}^{i+\frac{1}{2}} \end{array}\right)+\left(\begin{array}{c} 0 \\ v_{\delta\Omega~2}^i \end{array}\right)$
\item[5-] Set $\u^{i+1} = \u^{i+\frac{1}{2}}+\mathrm{P}_{\mathrm{div}\,0}\,\w^{i}$ 
\item[6-] Set $\boldsymbol\gamma_{\rm{res}}^{i+\frac{1}{2}} = \boldsymbol\gamma - \u^{i+1} - \v^i $
\item[7-] Set $\v^{i+\frac{1}{2}} = \v^{i}+\mathrm{Q}_{\mathrm{curl}\,0}\,\boldsymbol\gamma_{\rm{res}}^{i+\frac{1}{2}}$ with $\v^{i+\frac{1}{2}}=\sum_{jk}v_{jk}^{i+\frac{1}{2}}\Psi_{jk}^{\textrm{curl}}$
\item[8-] Compute $\v_{\delta \Omega}^{i+\frac{1}{2}}=\sum_{jk~{\rm s.t.}~\Psi_{jk\,|\partial\Omega}^{\textrm{div}}\neq 0}v_{jk}^{i+\frac{1}{2}}\Psi_{jk}^{\textrm{curl}}$
\item[9-] Set $ \w^{i+\frac{1}{2}} = \left(\begin{array}{c} 0 \\ -u_{{\delta \Omega}~2}^{i+1} \end{array}\right)+\left(\begin{array}{c} 0 \\ v_{{\delta \Omega}~2}^{i+\frac{1}{2}} \end{array}\right)$
\item[10-] Set $\v^{i+1} = \v^{i+\frac{1}{2}}-\mathrm{Q}_{\mathrm{curl}\,0}\,\w^{i+\frac{1}{2}}$ 
\end{itemize}
\end{itemize}

\bibliographystyle{aa}
\bibliography{wlens}
\end{document}